\documentclass[12pt]{article}
\usepackage{epsf}
\newcommand{\bmat}{\left(\begin{array}}
\newcommand{\emat}{\end{array}\right)}
\def\NPB{Nucl. Phys. B}
\def\PLB{Phys. Lett. B}
\def\PLB{Phys. Lett. B}

\def\yzero{\smash{\hbox{$y\kern-4pt\raise1pt\hbox{${}^\circ$}$}}}

\def\a{\alpha}
\def\b{\beta}
\def\l{\label}

\def\beq{\begin{equation}}
\def\eeq{\end{equation}}
\def\beqa{\begin{eqnarray}}
\def\eeqa{\end{eqnarray}}
\def\t{\times}

\def\-{\hphantom{-}}
\def\ov{\overline}
\def\s2{\frac{1}{\sqrt2}}

\def\beq{\begin{equation}}
\def\eeq{\end{equation}}
\def\beqa{\begin{eqnarray}}
\def\eeqa{\end{eqnarray}}

\def\IF{\relax{\rm I\kern-.18em F}}
\def\II{\relax{\rm I\kern-.18em I}}
\def\IP{\relax{\rm I\kern-.18em P}}
\def\IC{\relax\hbox{\kern.25em$\inbar\kern-.3em{\rm C}$}}
\def\IR{\relax{\rm I\kern-.18em R}}

\def\cp{{\cal P}}

\def\Dsl{\,\raise.15ex\hbox{/}\mkern-13.5mu D} %this one can be subscripted
\def\IZ{Z\kern-.4em  Z}

 \def\cp#1{\relax\ifmmode {\IP\kern-2pt{}_{#1}}\else $\IP\kern-2pt{}_{#1}$\=fi}
\catcode`\@=11
\newdimen\@rotdimen
\newbox\@rotbox

\def\@vspec#1{\special{ps:#1}}%  passes #1 verbatim to the output
\def\@rotstart#1{\@vspec{gsave currentpoint currentpoint translate
   #1 neg exch neg exch translate}}% #1 can be any origin-fixing transformation
\def\@rotfinish{\@vspec{currentpoint grestore moveto}}% gets back in synch
\def\@rotr#1{\@rotdimen=\ht#1\advance\@rotdimen by\dp#1%
   \hbox to\@rotdimen{\hskip\ht#1\vbox to\wd#1{\@rotstart{90 rotate}%
   \box#1\vss}\hss}\@rotfinish}
\def\@rotl#1{\@rotdimen=\ht#1\advance\@rotdimen by\dp#1%
   \hbox to\@rotdimen{\vbox to\wd#1{\vskip\wd#1\@rotstart{270 rotate}%
   \box#1\vss}\hss}\@rotfinish}%
\def\@rotu#1{\@rotdimen=\ht#1\advance\@rotdimen by\dp#1%
   \hbox to\wd#1{\hskip\wd#1\vbox to\@rotdimen{\vskip\@rotdimen
   \@rotstart{-1 dup scale}\box#1\vss}\hss}\@rotfinish}%
\def\@rotf#1{\hbox to\wd#1{\hskip\wd#1\@rotstart{-1 1 scale}%
   \box#1\hss}\@rotfinish}%
\def\rotate{\@ifnextchar[{\@rotate}{\@rotate[l]}}
\def\@rotate[#1]#2{\setbox\@rotbox=\hbox{#2}\@nameuse{@rot#1}\@rotbox}

\catcode`\@=12
\begin{document}

%----------------------------------------------------------------------%
%  numbering equations with section number
%----------------------------------------------------------------------%
\makeatletter \@addtoreset{equation}{section} \makeatother
\renewcommand{\theequation}{\thesection.\arabic{equation}}
%----------------------------------------------------------------------%
%  title page
%----------------------------------------------------------------------%
\pagestyle{empty}
%----------------------------------------------------------------------%
%  Resetting of counters
%----------------------------------------------------------------------%
%\setcounter{page}{0}
\pagestyle{empty}
\vspace{0.5in}
\rightline{FTUAM-02/35}
\rightline{IFT-UAM/CSIC-02-59}
\rightline{\today}
\vspace{2.0cm}
\setcounter{footnote}{0}

\begin{center}
\LARGE{
{\bf Deformed Intersecting D6-Brane GUTS and N=1 SUSY}}
\\[4mm]
%\medskip
{\large{ Christos ~Kokorelis \footnote{ Christos.Kokorelis@uam.es} }
\\[1mm]}
\normalsize{\em Departamento de F\'\i sica Te\'orica C-XI and 
Instituto de F\'\i sica 
Te\'orica C-XVI}
,\\[-0.3em]
{\em Universidad Aut\'onoma de Madrid, Cantoblanco, 28049, Madrid, Spain}
\end{center}
\vspace{1.0mm}

%%%%%%%%%%%%%%%%%%%%%%%%%%%%%%%%%%%%%%%

%\begin{center}
%\begin{minipage}[h]{14.5cm}
\begin{center}
{\small ABSTRACT}
\end{center}
We analyze the construction of non-supersymmetric
three generation
six-stack Pati-Salam (PS)
$SU(4)_C \times SU(2)_L \times SU(2)_R$ GUT classes of
models,
by localizing D6-branes
intersecting at angles in four dimensional
orientifolded toroidal
compactifications of type IIA. Special role in the models
is played by the presence of extra branes needed to satisfy the
RR tadpole cancellation conditions.
The models 
contain 
at low energy {\em exactly the Standard model}
 with no extra matter and/or
extra gauge group factors. They are build such that
they represent deformations around the quark and lepton
basic intersection number structure.
The models possess the same phenomenological
characteristics
of some recently discussed examples (PS-A , PS-I; PS-II
GUT classes; hep-th/0203187, hep-th/0209202;
hep-th/0210004) of
four and five stack
PS GUTS respectively. Namely,
there are no colour triplet couplings to mediate proton decay
 and proton is stable
 as baryon number is a gauged symmetry. 
 The mass relation
 $m_e = m_d$ at the GUT scale is recovered.
Even though more complicated, than in lower stack GUTS,
 the conditions of the non-anomalous U(1)'s to
 survive
 massless the generalized
 Green-Schwarz mechanism are solved consistently by the
 angle conditions coming from
 the presence of N=1 supersymmetric sectors involving
 the presence of {\em extra}
 branes and also required for the
 existence of a Majorana mass
 term for the right handed neutrinos.

\newpage

%----------------------------------------------------------------------%
%  Resetting of counters
%----------------------------------------------------------------------%
\setcounter{page}{1} \pagestyle{plain}
\renewcommand{\thefootnote}{\arabic{footnote}}
\setcounter{footnote}{0}
%----------------------------------------------------------------------%
%  Paper begins
%----------------------------------------------------------------------%

\section{Introduction}

In this work we 
present four dimensional (4D) three generation six
stack Pati-Salam (PS) like type of GUT classes of models
that break at low energy exactly to
the Standard Model (SM), 
$SU(3)_C \t SU(2)_L \t U(1)_Y$,
without any extra chiral fermions and/or extra gauge group 
factors (in the form of hidden sectors).  
These constructions are non-supersymmetric
and are centered around the PS
$SU(4)_C \times SU(2)_L \times SU(2)_R$ gauge
group.
These constructions have D6-branes intersecting each other
at angles,
in orientifolded compactifications of IIA theory in a
factorized six-tori,
with the O$_6$ orientifold planes on top of D6-branes
\cite{tessera, tessera2}.
\newline

The new classes of models have some
characteristic features that include :
\begin{itemize}

\item   The presented PS-III classes of  models,
are constructed with an initial
gauge group $U(4) \times
U(2) \times U(2) \times U(1)^3$ at the string
scale. At
the scale of
symmetry
breaking of the left-right symmetry 
 $M_{GUT}$, the initial symmetry group
breaks to the the standard model  
$SU(3)_C \times SU(2)_L \times U(1)_Y $
 augmented with two extra anomaly free $U(1)$ symmetries.
 The additional $U(1)$'s may break by gauge singlets generated
by imposing N=1 SUSY in particular sectors.

\item  A numbers of extra U(1)'s added to cancel the
RR tadpoles breaks by gauge singlets generated in imposing
N=1 SUSY on sectors involving the  extra U(1)'s.

\item  Neutrinos gets a mass 
 of the right order \footnote{In consistency with LSND oscillation 
experiments} 
from a see-saw mechanism of the Frogatt-Nielsen type.

\item  Proton stability is guaranteed due to the fact that baryon
number is an
unbroken gauged 
global symmetry surviving at low energies and no colour
triplet
couplings that could mediate
proton decay exist. 
A gauged baryon number is a general feature 
in D6-brane models \cite{louis2, kokos, kokoss, kokos1, kokos2, kokos3}

%\item
%The model uses small Higgs representations in the adjoint
%to break the PS symmetry,
%instead of using large
%Higgs representations \footnote{ E.g. 126 like in the
%standard $SO(10)$ models.}.

\end{itemize}

In the major problems of string 
theory the hierarchy of scale and 
particle masses after 
supersymmetry breaking is included. 
These issues have been explored
by explicitly
constructing semirealistic supersymmetric models 
of \footnote{
weakly coupled} $N=1$ orbifold compactifications of the 
heterotic string theories. One of the unsolved problems
is that the string scale which is of order $10^{18}$ GeV
is in clear disagreement with the `observed'
unification of
gauge
coupling constants in the MSSM of $10^{16}$ GeV.
The latter problem was not eventually solved 
even if
the discrepancy between the two high 
scales was attributed e.g. to
the presence of the 1-loop
string threshold corrections
to the gauge coupling constants \cite{dikomou}.

On the opposite side, in type I models, the string scale 
is a free parameter and thus may be lowered in the
TeV region \cite{antoba} suggesting 
that non-SUSY models with a TeV string scale 
is a possibility.
In this content new constructions have
appeared in type I
string vacuum background which 
construct even generation four dimensional non-supersymmetric models
using intersecting branes.  The new constructions were made possible
 by turning on
background fluxes on D9 branes 
on a type I background \footnote{
In the T-dual backgrounds these constructions are represented by
D6 branes wrapping 3-cycles on a dual six dimensional torus
and intersecting each other at certain angles }.
Thus open string models were constructed that
break supersymmetry on the 
brane and give chiral fermions 
with an even number of 
generations \cite{tessera}.
In these models the fermions get localized
in the
intersections between
branes \cite{bele}. 
The
introduction of a quantized background NS-NS B
field \cite{eksi1,eksi2,eksi3}, that makes the tori tilted,
subsequently
give rise to 
semirealistic models with 
three generations \cite{tessera2}.
The latter backgrounds
are T-dual to models with magnetic
deformations \cite{carlo}.
In \cite{louis2} the first examples of four D6-brane stack 
models which have only the SM at 
low energy, in the language
of D6-branes intersecting at angles on an orientifolded 
$T^6$ torus, were constructed.  
In this
construction, as in all models from the same backgrounds,
 the proton is stable since the baryon number
is a gauged
$U(1)$ global symmetry.
A special feature of these models is that the neutrinos
can only get Dirac mass. These models have been
generalized to models with
five stacks \cite{kokos} and six stacks 
of D6-branes at the string scale \cite{kokos2}.
The models appearing in \cite{kokos} \cite{kokos2} are build as
novel deformations of the QCD intersection numbers, namely
they are build around the left and right handed quarks
intersection numbers. They hold exactly the same
phenomenological properties of \cite{louis2}. Its is important to stress 
here that contrary to their four stack counterparts \cite{louis2} they
have unique special features, since by demanding
the presence of $N=1$ supersymmetric sectors, we are
able to break the extra, beyond the SM gauge group,
$U(1)$'s, and thus predicting the unique existence of one
supersymmetric partner of the right neutrino or
two supersymmetric partners of the right neutrinos in
the five and the six stack SM's respectively.

In a parall development, in \cite{kokos1} we presented the first
examples of GUT models in a string theory context, and 
in the context of intersecting
branes, that break completely to the SM at
low energies. These models predict the unique existence
of light weak fermion doublets with energy between 
 $M_Z$ - 246 GeV, thus can be directly tested at
present of future accelerators. 
The constructions of four D6-brane stack of \cite{kokos1}
were studied further in \cite{kokos2} in relation to other four stack 
deformations, and by extending these constructions to five stacks 
of D6-brane stacks in \cite{kokos3}.
In this work we will discuss the six extensions of \cite{kokos1, kokos2, 
kokos3}.

We note that apart from D6 models with exactly the SM at low
energy just mentioned, there intersecting
D5-branes have been studied with only the SM at low energy \cite{D5, D51}. 
In the latter models \cite{D51} there are special classes of theories,
again appearing as novel deformations of the QCD intersection 
number structure, which have 
not only the SM at low energy but exactly the same
low energy effective
theory including fermion and scalar spectrum.

For additional work , in the context of
intersecting branes see
\cite{luis1, uran, fors, cala, uran1, cim, cim1, maria, dimi} 
\footnote{For some other constructions close to the SM but not based
on a particular string constructions see \cite{antokt1}.}.

The paper is organized as follows: 
 In section two 
 we describe the general spectra and tadpole rules for
 building chiral GUT models in orientifolded $T^6$
 compactifications. In section 3, 
we discuss the basic
 fermion and scalar structure of the present PS-III class
 of models.
 In section 4,
 we
discuss the parametric solutions to the
RR tadpole cancellation conditions as well giving the general 
characteristics of the PS-III GUTS.
In section 5 we analyze the cancellation of $U(1)$
anomalies in
the presence of a 
generalized 
Green-Schwarz (GS) mechanism
and extra $U(1)$ branes.
In section 6, we discuss the conditions for the
absence of tachyons, the angle structure between the branes 
and its role in describing the
 Higgs sector
of the models.     
In subsection 7.1 we discuss the importance of creating
sectors preserving N=1 SUSY for the realization of the
see-saw mechanism and its relation to the generalized 
Green-Schwarz mechanism.
In subsection 7.2 we discuss the role of
extra U(1) branes in creating scalar
singlets.
In subsection 7.3 we analyze the breaking of the
surviving the Green-Schwarz mechanism massless $U(1)$'s. 
In subsection 8.1 we analyze the structure of GUT
Yukawa
couplings in intersecting braneworlds and 
discuss the problem of
neutrino masses.
In subsection 8.2
we exhibit that all
additional exotic fermions beyond those of SM present in the
models may become
massive
and disappear from the low energy $M_Z$ spectrum.
We present our conclusions in Section 9.  
Appendix I, includes the conditions for the
absence of tachyonic modes
in the spectrum of the PS-III class of models discussed in this work.

\section{\em Tadpole structure and spectrum rules}

Let us now describe the construction of the PS classes of models. It is 
based on  
type I string with D9-branes compactified on a six-dimensional
orientifolded torus $T^6$, 
where internal background 
gauge fluxes on the branes are turned on \cite{bachas, tessera, tessera2}. 
By performing a T-duality transformation on the $x^4$, $x^6$, $x^8$, 
directions the D9-branes with fluxes are translated into D6-branes 
intersecting at 
angles. 
We assume that the D$6_a$-branes are wrapping 1-cycles 
$(n^i_a, m^i_a)$ along each of the $T^2$
torus of the factorized $T^6$ torus, namely 
$T^6 = T^2 \times T^2 \times T^2$.

In order to build a PS model with a minimal Higgs
structure we consider
six stacks of D6-branes giving rise to their world-volume to an initial
gauge group $U(4) \times U(2) \times U(2) \times U(1) \times U(1) \times U(1)$
gauge symmetry at the string
scale.
In addition, we consider the addition of NS B-flux, 
such that the tori are not orthogonal,
avoiding in this way an even number of families, 
and leading to effective tilted wrapping numbers, 
\beq
(n^i, m ={\tilde m}^i + {n^i}/2);\; n,\;{\tilde m}\;\in\;Z,
\label{na2}
\eeq
that allows semi-integer values for the m-numbers.  
\newline
In the presence of $\Omega {\cal R}$ symmetry, 
where $\Omega$ is the worldvolume 
parity and $\cal R$ is the reflection on the T-dualized coordinates,
\beq
T (\Omega )T^{-1}= \Omega {\cal R},
\label{dual}
\eeq
and thus each D$6_a$-brane
1-cycle, must have its $\Omega {\cal R}$ partner $(n^i_a, -m^i_a)$. 

Chiral fermions are obtained by stretched open strings between
intersecting D6-branes \cite{bele}. 
Also the chiral spectrum of the models may be obtained 
after solving simultaneously 
the intersection 
constraints coming from the existence of the different sectors taking into 
account the RR
tadpole cancellation conditions.

There are a number of different sectors, 
contributing to the chiral spectrum.
Denoting the action of 
$\Omega R$ on a sector $\a, \b$, by ${\a}^{\star},
{\b}^{\star}$,
respectively the possible sectors are:

\begin{itemize}
 
\item The $\a \b + \b \a$ sector: involves open strings stretching
between the
D$6_{\a}$ and D$6_{\b}$ branes. Under the $\Omega R$ symmetry 
this sector is mapped to its image, ${\a}^{\star} {\b}^{\star}
+ {\b}^{\star} {\a}^{\star}$ sector.
The number, $I_{\a\b}$, of chiral fermions in this sector, 
transforms in the bifundamental representation
$(N_{\a}, {\bar N}_{\a})$ of $U(N_{\a}) \times U(N_{\b})$,
being
\beq
I_{\a\b} = ( n_{\a}^1 m_{\b}^1 - m_{\a}^1 n_{\b}^1)( n_{\a}^2 m_{\b}^2 -
m_{\a}^2 n_{\b}^2 )
(n_{\a}^3 m_{\b}^3 - m_{\a}^3 n_{\b}^3),
\label{ena3}
\eeq
 where $I_{\a\b}$
is the intersection number of the wrapped three cycles.
Note that with the sign of
 $I_{\a\b}$ we denote the chirality of the fermions,
$I_{\a\b} > 0$ being those of left handed fermions.
Negative multiplicity correspond to opposite chirality.

\item The $\a\a$ sector : it involves open strings stretching on a single 
stack of 
D$6_{\a}$ branes.  Under the $\Omega R$ symmetry 
this sector is mapped to its image ${\a}^{\star}{\a}^{\star}$.
 This sector contain ${\cal N}=4$ super Yang-Mills and if
 it exists
SO(N), SP(N) groups in principle may appear. 
This sector is of no importance to us as we will be dealing with
 unitary groups.

\item The ${\a} {\b}^{\star} + {\b}^{\star} {\a}$ sector :
In this sector which under the $\Omega R$ symmetry
transforms to itself, chiral fermions transform
into the $(N_{\a}, N_{\b})$
representation with multiplicity given by
\beq
I_{{\a}{\b}^{\star}} = -( n_{\a}^1 m_{\b}^1 + m_{\a}^1 n_{\b}^1)
( n_{\a}^2 m_{\b}^2 + m_{\a}^2 n_{\b}^2 )
(n_{\a}^3 m_{\b}^3 + m_{\a}^3 n_{\b}^3).
\label{ena31}
\eeq

\item the ${\a} {\a}^{\star}$ sector :
under the $\Omega R$ symmetry is
transformed to itself. In this sector the
invariant intersections
give 8$m_{\a}^1 m_{\a}^2 m_{\a}^3$ fermions in the
antisymmetric representation
and the non-invariant intersections that come in pairs
provide 
4$ m_{\a}^1 m_{\a}^2 m_{\a}^3 (n_{\a}^1 n_{\a}^2 n_{\a}^3 -1)$ additional 
fermions in the symmetric and 
antisymmetric representation of the $U(N_{\a})$ gauge
group.

\end{itemize}

Additionally any vacuum derived from the previous
intersection number constraints of the
chiral spectrum 
is subject to constraints coming from RR tadpole
cancellation
conditions \cite{tessera}. That is equivalent to
cancellation of
D6-branes charges \footnote{Taken together with their
orientifold images $(n_a^i, - m_a^i)$  wrapping
on three cycles of homology
class $[\Pi_{\alpha^{\star}}]$.}, wrapping on three cycles with
homology $[\Pi_a]$ and O6-plane 7-form
charges wrapping on 3-cycles with homology $[\Pi_{O_6}]$.
Explicitly, 
the RR tadpole cancellation  conditions expressed
in terms of cancellations of RR charges in homology, obey :
\beq
\sum_a N_a [\Pi_a]+\sum_{\alpha^{\star}} 
N_{\alpha^{\star}}[\Pi_{\alpha^{\star}}] -32
[\Pi_{O_6}]=0.
\label{homology}
\eeq  
or
\beqa
\sum_a N_a n_a^1 n_a^2 n_a^3 =\ 16,\nonumber\\
\sum_a N_a m_a^1 m_a^2 n_a^3 =\ 0,\nonumber\\
\sum_a N_a m_a^1 n_a^2 m_a^3 =\ 0,\nonumber\\
\sum_a N_a n_a^1 m_a^2 m_a^3 =\ 0.
\label{na1}
\eeqa
That forces absence of non-abelian gauge anomalies.

A six-stack brane configuration with minimal PS particle
content may be 
obtained by considering six stacks of branes yielding an initial
$U(4)_a \times U(2)_b \times U(2)_c \times U(1)_d \times U(1)_e \times U(1)_f$.
In this
case the equivalent
gauge group is an $SU(4)_a \times SU(2)_b \times SU(2)_b 
\times U(1)_a \times U(1)_b 
\times U(1)_c \times U(1)_d \times U(1)_e \times U(1)_f$.
Thus, in the first instance,
we identify, without loss of 
generality, $SU(4)_a$ as the $SU(4)_c$ colour group 
that its breaking induces 
the $SU(3)$ colour group of strong interactions,
the $SU(2)_b$ with
$SU(2)_L$ of weak interactions and $SU(2)_c$ with
the $SU(2)_R$  of left-right symmetric PS models.

\section{\em The basic fermion  structure}

In this section we will describe the basic
characteristics of the GUT models that we will
analyze in this work.
The models are three family non-supersymmetric
GUT model with the left-right
symmetric PS model structure \cite{pati}
$SU(4)_C \times SU(2)_L \times SU(2)_R$.
The open string background from which the models will be
derived are intersecting D6-branes wrapping
on 3-cycles of
decomposable toroidal ($T^6$) orientifolds of type IIA
in four dimensions \cite{tessera, tessera2}.
 
The three generations of quark and lepton 
fields are accommodated into the following
representations :
\beqa
F_L &=& (4, {\bar 2}, 1) =\ q(3, {\bar 2}, \frac{1}{6})
+\ l(1, {\bar 2}, -\frac{1}{2})
\equiv\ (\ u,\ d,\ l), \nonumber\\
{\bar F}_R &=& ({\bar 4}, 1,  2) =\ {u}^c({\bar 3}, 1, - \frac{2}{3}) +
{d}^c({\bar 3}, 1, \frac{1}{3}) + {e}^c(1, 1, 1) + {N}^c(1, 1, 0) 
\equiv ( {u}^c,  {d}^c,   {l}^c),\nonumber\\ 
\label{na3}
\eeqa
where the quantum numbers on the right hand side of 
(\ref{na3})
are with respect to the decomposition of the $SU(4)_C \times SU(2)_L \times 
SU(2)_R$ under the $SU(3)_C \times SU(2)_L \times U(1)_Y$ gauge group and
${l}=(\nu,e) $ is the standard left handed lepton doublet
while
 ${l}^c=({N}^c, e^c)$ are the right handed leptons.
The assignment of the accommodation of the quarks and leptons into
the representations $F_L + {\bar F}_R$ is the one appearing in the
spinorial decomposition of the $16$ representation
of $SO(10)$ under the PS gauge group.
\newline
Also present are the fermions 
\beq
\chi_L =\ ( 1,  {\bar 2}, 1),\ \chi_R =\ (1, 1, {\bar 2}).
\label{useful}
\eeq
These fermions are a general prediction
of left-right symmetric theories as their existence
follows from RR tadpole cancellation conditions. 
\newline
The symmetry breaking of the left-right PS symmetry
at the $M_{GUT}$ scale,  in principle
as high as the
string scale,  
proceeds through the representations of
the set of Higgs fields,
\beq
H_1 =\ ({\bar 4}, 1, {\bar 2}), \
H_2 =\ ( 4, 1, 2),
\label{useful1}
\eeq
 where,
\beqa
H_1 = ({\bar 4}, 1, {\bar 2}) = {u}_H({\bar 3}, 1, \frac{2}{3}) +
{d}_H({\bar 3}, 1, -\frac{1}{3}) + {e}_H(1, 1, -1)+
{\nu}_H(1, 1, 0).
\label{higgs1}
\eeqa
The electroweak symmetry breaking is achieved 
through bi-doublet Higgs fields $h_i$ $i=3, 4$
in the representations
\beq
h_3 =\ (1, {\bar 2}, 2),\ h_4 =\ (1, 2, {\bar 2}) \ .
\label{additi1}
\eeq
Because of the imposition of N=1 SUSY on some open string
sectors, there are also present
 the massless scalar superpartners of the quarks, leptons and
antiparticles 
 \beq
{\bar F}_R^H = ({\bar 4}, 1,  2) = {u}^c_H({\bar 3}, 1, -\frac{4}{6})+
{d}^c_H({\bar 3}, 1, \frac{1}{3})+ {e}^c_H(1, 1, 1) + 
{N}^c_H(1, 1, 0) 
\equiv ({u}^c_H, {d}^c_H,  {l}^c_H).\nonumber\\
\label{na368}
\eeq
The latter fields \footnote{
are replicas
of the fermion fields appearing in the intersection 
$ac$ and they receive a vev} characterize all vacua
coming from these type IIA orientifolded tori 
constructions is the replication of massless fermion spectrum
by an equal number of massive particles in the
same representations and with the same quantum numbers.
\newline
Also, a number of charged exotic
fermion fields, which may receive a string scale mass,
appear
\beq
6(6, 1, 1),\;\; \ 6({\bar 10}, 1, 1).
\label{beg1}
\eeq
In addition the following gauge singlet fermion
fields appear : $s_R^1$, $s_R^2$, $s_R^3$.

The complete accommodation of the fermion structure of
the PS-III classes of models can be seen
in table  (\ref{spectrum8}).

\begin{table}[htb] \footnotesize
\renewcommand{\arraystretch}{1.5}
\begin{center}                          
\begin{tabular}{|c|c||c|c||c||c|c|c|c|}
\hline
Fields &Intersection  & $\bullet$ $SU(4)_C \times SU(2)_L \times SU(2)_R$
 $\bullet$&
$Q_a$ & $Q_b$ & $Q_c$ & $Q_d$ & $Q_e$ & $Q_f$\\
\hline
 $F_L$& $I_{ab^{\ast}}=3$ &
$3 \times (4,  2, 1)$ & $1$ & $1$ & $0$ &$0$ &$0$ &$0$\\
 ${\bar F}_R$  &$I_{a c}=-3 $ & $3 \times ({\ov 4}, 1, 2)$ &
$-1$ & $0$ & $1$ & $0$ & $0$ & $0$\\\hline
 $\chi_L^1$& $I_{bd} = -6$ &  $6 \times (1, {\ov 2}, 1)$ &
$0$ & $-1$ & $0$ & $1$ & $0$ & $0$\\    
 $\chi_R^1$& $I_{cd^{\star}} = -6$ &
 $6 \times (1, 1, {\ov 2})$ &
$0$ & $0$ & $-1$ & $-1$ & $0$ & $0$\\
 $\chi_L^2$& $I_{be} = -4$
 &  $4 \times (1, {\ov 2}, 1)$ &
$0$ & $-1$ & $0$ &$0$ & $1$ & $0$\\    
 $\chi_R^2$ & $I_{ce^{\ast}} = -4$ &  $4 \times (1, 1, {\ov 2})$ &
$0$ & $0$ & $-1$ & $0$ &$-1$ & $0$\\
 $\chi_L^3$ & $I_{bf} = -2$ &  $2 \times (1, {\bar 2}, 1)$ &
$0$ & $-1$ & $0$ & $0$ & $0$ & $1$\\
 $\chi_R^3$ & $I_{cf^{\star}} = -2$ &
 $2 \times (1, 1, {\bar 2})$ &
$0$ & $0$ & $-1$ & $0$ & $0$ & $-1$\\\hline
 $\omega_R$ & $I_{aa^{\ast}}$ &  $6 \b^2
 \times (6, 1, 1)$ & $2$ & $0$ & $0$ &$0$ & $0$ & $0$\\
 $y_R$& $I_{aa^{\ast}}$ & $6  \b^2  \times ({\bar 10}, 1, 1)$ &
$-2$ & $0$ & $0$ &$0$ &$0$ & $0$\\
\hline
 $s_R^1$ & $I_{dd^{\ast}}$ &  $12 \b^2
 \times (1, 1, 1)$ & $0$ & $0$ & $0$ & $-2$ &$0$ & $0$\\
 $s_R^2$ & $I_{ee^{\ast}}$ & $8  \b^2
 \times (1, 1, 1)$ & $0$ & $0$ & $0$ &$0$ &$-2$ & $0$\\
 $s_R^3$ & $I_{ff^{\ast}}$ & $8  \b^2
 \times (1, 1, 1)$ & $0$ & $0$ & $0$ &$0$ &$0$ & $-2$\\
\hline
\end{tabular}
\end{center}
\caption{\small
 Fermionic spectrum of the $SU(4)_C \times
SU(2)_L \times SU(2)_R$, PS-III class of models together with $U(1)$
charges. We note that at energies of order $M_z$ only
the Standard model survives.
\label{spectrum8}}
\end{table}

\section{\em Tadpole cancellation for PS-III
classes of GUTS}

The following comments will be necessary to understand
the analysis performed in the following sections
of the PS-III classes of GUTS :
\newline
a) A proper formulation of a GUT model requires the
realization of certain couplings necessary to e.g.
for giving masses to right handed neutrinos or
make massive non-observed particles.
In intersecting brane worlds based on intersecting
D6-branes this becomes much easier by demanding
that some open string sectors preserve some
supersymmetry. Thus some massive
fields will be
pulled out from the massive spectrum and become massless.
Thus a Majorana mass term
for the right
handed neutrinos is realized if the sector
$ac$ preserves $N=1$ SUSY.
As an immediate effect the previously massive
${\bar F}_R^H$ scalar appears.
\newline
b)
The intersection numbers, in table (\ref{spectrum8}),  
of the fermions $F_L + {\bar F}_R$ are chosen 
such that $I_{ac} = - 3$, $I_{ab^{\star}} = 3$. Here, $-3$ denotes opposite 
chirality to that of a left handed fermion. 
The choice of additional fermion representations
$(1, {\bar 2} ,1)$,
$(1, 1, {\bar 2})$ is imposed to us by the RR tadpole cancellation conditions
that are equivalent to
$SU(N_a)$ gauge
anomaly cancellation, in this case of $SU(2)_L$,
$SU(2)_R$ gauge anomalies,
\beq
\sum_i I_{ia} N_a = 0,\;\;a = L, R.
\label{ena4}
\eeq
c)
In the present classes of models
representations
of scalar sextets $(6, 1, 1)$ fields,
that appear in attempts to construct realistic
4D $N=1$ PS
heterotic models
from
the fermionic formulation \cite{antoI},
even 
through heterotic fermionic models where those 
representations are lacking 
exist \cite{giapo}, do not appear.
These representations were imposed in
attempts to produce a
realistic PS model in order to save the models from fast 
proton decay.
Thus fast proton decay was avoided by making 
the mediating 
$d_H$ triplets of (\ref{higgs1}) superheavy 
and of order of the $SU(2)_R$ breaking scale
through their couplings to the sextets.
In the PS-III models
baryon number is a gauged global symmetry, thus proton is 
stable. Hence there is no need to introduce proton decay saving sextets.
\newline
Moreover in the PS-III GUTS, there is no problem of having
$d_H$ becoming light
enough and inducing proton decay, as the only way this could
happen, is through the existence of $d_H$
coupling of sextets
to quarks and leptons. However, this coupling is forbidden by
the symmetries of the models.
\newline
d) The mixed anomalies $A_{ij}$ of the six \footnote{We examine for 
convenience the case of two added extra U(1)'s.}
surplus $U(1)$'s
with the non-abelian gauge groups $SU(N_a)$ of the theory
cancel through a generalized GS mechanism \cite{sa, louis2},
involving
 close string modes couplings to worldsheet gauge fields.
 Crucial for the RR tadpole
 cancellation is the presence of $N_h$ extra branes.
 Contrary,
 of what
 was found in D6-brane models 
  with exactly the SM at low energy,
  and a Standard-like structure at the string
  scale \cite{louis2, kokos, kokos2} where
  the extra branes have no intersection with the 
branes \footnote{A similar phenomenon
 appears in  
 intersecting D5-brane models on a $T^4 \times C/Z_N$, 
with exactly the SM at
 low energy
  and a Standard-like structure at the string
  scale \cite{D5, D51}.},
  in the present PS-III GUT models there is a
  non-trivial intersection of the
extra branes with the branes $a$, $b$, $c$.
  As a result, this becomes a new singlet generation
  mechanism after imposing $N=1$ SUSY between $U(1)$
   leptonic (the $d$, $e$,  $f$ branes) and the
  $U(1)$ extra branes.
  Also, contrary to the SM's of
  \cite{louis2, kokos, kokoss, D5, D51} the extra branes
  do not form a $U(N_h)$ gauge group but rather a
   $U(1)^{N_1} \times U(1)^{N_2} \cdots U(1)^{N_h}$ one,
   where $ N_1 = N_2 = \cdots = N_h = 1$.
\newline
e)
We don't impose the constraint  
\beqa
{\Pi}_{i=1}^3 m^i&=&0.
\label{req1}
\eeqa                  
As a result chiral
fermions appear from the
$aa^{\ast}$, $dd^{\ast}$, $ee^{\ast}$, $ff^{\ast}$, sectors
 with corresponding
fermions $\omega_L$, $y_R$; $s_R^1$; $s_R^2$; $s_R^3$.\newline
f)
The PS left-right symmetry is being broken
at $M_{GUT}$. As there is no constraint
from first principles for $M_{GUT}$ we shall take it equal to
the string scale. The
surviving gauge symmetry 
is that of the SM augmented by five anomaly free $U(1)$
symmetries,
including the two chosen added extra U(1) branes,
surviving the Green-Schwarz mechanism.  
The breaking of the latter $U(1)$ symmetries
will be facilitated by 
having the $dd^{\star}$, $ee^{\star}$, $ff^{\star}$, $dh$,
$dh^{\star}$, $eh$,$eh^{\star}$,$fh$, $fh^{\star}$
sectors preserving N=1 SUSY \footnote{We denoted by $h$ the presence of 
extra U(1) branes.}.
Thus singlets scalars will be localized in these
intersections, that are
superpartners of the corresponding fermions.
\newline
g) The third tori is permanently tilded. Also
b-brane is parallel to the c-brane (and the $a$ D6-brane  is 
parallel to the $d$, $e$, $f$ D6-branes).
The  
cancellation of the RR tadpole constraints
is solved from multiparameter sets of solutions.
They are given
in table (\ref{spectruma101}). Their satisfaction needs a
number of extra branes, positioned at
$(1/\beta_1, 0) (1/\beta_2, 0),(2,0)$, 
 having non-trivial intersection
numbers with the a, d, e, f branes and thus creating extra
fermions which finally become massive by arranging
for some sectors to have N=1 SUSY.

\begin{table}[htb]\footnotesize
\renewcommand{\arraystretch}{1.7}
\begin{center}
\begin{tabular}{||c||c|c|c||}
\hline
\hline
$N_i$ & $(n_i^1, m_i^1)$ & $(n_i^2, m_i^2)$ & $(n_i^3, m_i^3)$\\
\hline\hline
 $N_a=4$ & $(0, \epsilon)$  &
$(n_a^2, 3 \epsilon {\tilde \epsilon}\b_2)$ & $(1, {\tilde \epsilon}/2)$  \\
\hline
$N_b=2$  & $(-1, \epsilon m_b^1 )$ & $(1/\beta_2, 0)$ &
$(1, {\tilde \epsilon}/2)$ \\
\hline
$N_c=2$ & $(1, \epsilon m_c^1 )$ &   $(1/\beta_2, 0)$  & 
$(1, -{\tilde \epsilon}/2)$ \\    
\hline
$N_d=1$ & $(0, \epsilon)$ &  $(n_d^2, -3 \epsilon
{\tilde \epsilon}\b_2)$
  & $(2, -{\tilde \epsilon})$  \\\hline
$N_e=1$ & $(0, \epsilon)$ &  $(n_e^2, -2 \epsilon
{\tilde \epsilon}\b_2)$
  & $(2, -{\tilde \epsilon})$  \\\hline
$N_f=1$ & $(0, \epsilon)$ &  $(n_f^2, -2 \epsilon
{\tilde \epsilon}\b_2)$
  & $(1, -\frac{1}{2}{\tilde \epsilon})$  \\\hline
$1$& $(1/\beta_1, 0)$ &  $(1/\beta_2, 0)$
  & $(2, 0)$  \\\hline
$\vdots$& $\vdots$ &  $\vdots$
  & $\vdots$  \\\hline
$N_h$ & $(1/\beta_1, 0)$ &  $(1/\beta_2, 0)$
  & $(2, 0)$  \\\hline   
\end{tabular}
\end{center}
\caption{\small
RR tadpole solutions for PS-III classes of GUT models
with six stacks of intersecting
D6-branes giving rise to the 
fermionic spectrum and the SM,
$SU(3)_C \times SU(2)_L \times U(1)_Y$, gauge group at low energies.
The wrappings 
depend on four integer parameters, 
$n_a^2$, $n_d^2$, $n_e^2$, $n_f^2$, the NS-background $\beta_i$ and the 
phase parameters $\epsilon = {\tilde \epsilon }= \pm 1$. 
Also there is an additional dependence on the two wrapping
numbers, integer of half integer,
$m_b^1$, $m_c^1$. Note also that the presence of the
$N_h$ extra U(1) branes.}
\label{spectruma101}
\end{table}

The first tadpole condition in (\ref{na1}) 
depends on the number of extra branes 
\beq
N_h \frac{2}{\beta_1 \beta_2} = 16.
\eeq
Also the third tadpole condition reads :
\beq
(2 n_a^2 - n_d^2 - n_e^2-\frac{1}{2}n_f^2)
+ \frac{1}{\beta_2}(m_b^1 -m_c^1) =\ 0.
\label{ena11}
\eeq

h) the hypercharge operator 
 is defined as usual in this classes of GUT
 models( see also \cite{kokos1}) as
 a linear combination
of the three diagonal generators of the $SU(4)$, $SU(2)_L$, $SU(2)_R$ groups:
\beq
Y = \frac{1}{2}T_{3R}+ \frac{1}{2}T_{B-L},\;T_{3R}=diag(1,-1),\;
T_{B-L}=diag(\frac{1}{3},\frac{1}{3},\frac{1}{3}, -1). 
\label{hyper12}
\eeq 
Also,
\beqa
Q & = & Y   +\ \frac{1}{2}T_{3L} \ .
\label{hye1}
\eeqa

\section{\em Cancellation of U(1) Anomalies}

In order to saw that the classes of models described by the solutions to the
RR tadpoles of table (2) break at low energy to the SM we have first to
exhibit that the massless additional U(1)'s originally present in 
the models in the
string scale receive a mass and disappear from the low energy spectrum.

The mixed anomalies $A_{ij}$ of the six $U(1)$'s
with the non-Abelian gauge groups are given by
\beq
A_{ij}= \frac{1}{2}(I_{ij} - I_{i{j^{\star}}})N_i.
\label{ena9}
\eeq
%
%Analyzing the mixed anomalies 
%of the extra $U(1)$'s with the non-abelian gauge groups $SU(4)_c$, 
%$SU(2)_R$, $SU(2)_L$ we can see that there are three anomaly free combinations
%$Q_b - Q_c$, $Q_a + Q_d - Q_e$ and $Q_a + 4 Q_d + 5 Q_e$.
In the orientifolded type IIA toroidal models the gauge anomaly 
cancellation \cite{sa} is achieved through a 
generalized GS
mechanism \cite{louis2} that makes use of the 10-dimensional RR 
gauge fields
$C_2$ and $C_6$ and gives at four dimensions \footnote{ 
Note that gravitational anomalies cancel since D6-branes never 
intersect O6-planes.
 }
the couplings to gauge fields 
 \beqa
N_a m_a^1 m_a^2 m_a^3 \int_{M_4} B_2^o \wedge F_a &;& n_b^1 n_b^2 n_b^3
 \int_{M_4}
C^o \wedge F_b\wedge F_b,\\
N_a  n^J n^K m^I \int_{M_4}B_2^I\wedge F_a&;&n_b^I m_b^J m_b^K \int_{M_4}
C^I \wedge F_b\wedge F_b\;,
\label{ena66}
\eeqa
where
$C_2\equiv B_2^o$ and $B_2^I \equiv \int_{(T^2)^J \times (T^2)^K} C_6 $
with $I=1, 2, 3$ and $I \neq J \neq  K $. Notice the four dimensional duals
of $B_2^o,\ B_2^I$ :
\beqa
C^o \equiv \int_{(T^2)^1 \times (T^2)^2 \times (T^2)^3} C_6&;C^I \equiv
\int_{(T^2)^I} C_2, 
\label{ena7}
\eeqa
where $dC^o =-{\star} dB_2^o,\; dC^I=-{\star} dB_2^I$.

The triangle anomalies (\ref{ena9}) cancel from the existence of the
string amplitude involved in the GS mechanism \cite{sa} in four 
dimensions \cite{louis2}. 
The latter amplitude, where the $U(1)_a$ gauge field couples to one
of the propagating
$B_2$ fields, coupled to dual scalars, that couple in turn to
two $SU(N)$ gauge bosons, is 
proportional \cite{louis2} to
\beq
-N_a  m^1_a m^2_a m^3_a n^1_b n^2_b n^3_b -
N_a \sum_I n^I_a n^J_a n^K_b m^I_a m^J_b m^K_b\; ,
I \neq J, K 
\label{ena8}
\eeq

We make the minimal choice
\beq
\beta_1 = \beta_2 =  1/2
\eeq
that requires two extra D6 branes. \newline
In this case
the structure of U(1) couplings reads :
\beqa
B_2^3 \wedge 
[-\frac{{\tilde \epsilon}}{\beta^2}]
[(F^b + F^c)],&\nonumber\\
B_2^1 \wedge [\epsilon] 
[ 4 n_a^2 \ F^a +
2\frac{m_b^1}{\beta^2} \ F^b + 2\frac{ m_c^1}{\beta^2}
\ F^c + 2 n_d^2  F^d +
2 n_e^2 F^e + n_f^2 F^f],&\nonumber\\
B_2^o  \wedge (\beta^2) \left(   6  F^a + 3 F^d
+ 2 F^e + F^f \right) .&
\label{PSIIb}
\eeqa
As can be seen from (\ref{PSIIb}) two anomalous
combinations of $U(1)$'s, e.g.
 $6 F^a +3 F^d + 2F^e + F^f$, $(F^b + F^c)$
 become massive through their couplings to RR
 fields $B_2^o$, $B_2^3$. In addition,  
there is an anomaly free
 model dependent U(1) which is getting massive via
 its coupling to the RR field $B_2^1$. 
In addition, there are three non-anomalous $U(1)$'s, that may broken 
by the vevs of singlet scalars generated either,
by imposing N=1 SUSY on sectors of the form $dd^{*}$, $ee^{*}$, $ff^{*}$, or
by sectors involving the presence of the extra branes (see later discussion). 
They are :
\beqa
U(1)^{(4)} =\ (Q^b - Q^c) + (Q^a - Q^d - Q^e - Q^f),
&\nonumber\\
U(1)^{(5)} =\ \frac{1}{3} Q^a - \frac{15}{9}Q^d
+ Q^e + Q^f\nonumber\\
U(1)^{(6)} =\ 2 Q^a + Q^d -16 Q^e + 17 Q^f\ .
\label{PSIIab1}
\eeqa
Crucial for the satisfaction of RR tadpoles is the
addition of $N_h$ hidden branes.
We note that for simplicity that when
$\beta^1 = \beta^2 = 1/2$ , $N_h = 2$.
In this case, we simply have to add these extra
U(1)'s to the bunch of (\ref{PSIIab1}),
\beqa
U(1)^{(7)} =\ F^{{\hat h}_1}, & 
U(1)^{(8)} =\ F^{{\hat h}_2}. 
\label{PSIIab12}
\eeqa
We note that the model independent U(1)'s (\ref{PSIIab1}),
survive massless the presence of 
the generalized Green-Shcwarz mechanism imposed by the existence of the 
couplings (\ref{PSIIb}), as long as
\beqa
12 n_a^2 -\ 30 n_d^2 +\ 18 n_e^2+ 9 n_f^2 =\ 0, \nonumber\\
8 n_a^2 +\ 2 n_d^2 -\ 32 n_e^2 + 17 n_f^2 =\ 0
\label{constr11}
\eeqa
They can broken by the existence of singlets 
generated either,
by imposing N=1 SUSY on sectors of the form $dd^{*}$, $ee^{*}$,
$ff^{*}$, or
on sectors involving the presence of the extra branes
needed to satisfy the RR tadpole cancellation conditions.
The reader should notice that the conditions for demanding that some 
sectors respect N=1 SUSY, that in turn guarantee the existence
of a Majorana coupling for right handed neutrinos as well
creating singlets
necessary to break the U(1)'s (\ref{PSIIab1}), solve the condition
(\ref{constr11}). They are analyzed in section (7.1).

Also the couplings of the dual scalars $C^I$ of $B_2^I$,
required to cancel
the mixed anomalies of the $U(1)$- 
non-abelian $SU(N_a)$ anomalies, appear as :
\beqa
C^o \wedge   [-\frac{1}{\beta^2}][(F^b \wedge F^b)
- (F^c \wedge F^c) - (\frac{2}{\beta^1})(F^{h^1} \wedge F^{h^1} 
+ F^{h^2} \wedge F^{h^2}) ], &\nonumber\\
C^2 \wedge [\frac{{\epsilon}{\tilde \epsilon}}{2}]
[ n_a^2 (F^a \wedge
F^a)  +   \frac{m_b^1}{\beta^2} (F^b \wedge 
F^b) -  \frac{ m_c^1}{\beta^2} (F^c \wedge F^c) 
- 2n_d^2 (F^d \wedge F^d) &\nonumber\\
- 2n_e^2 (F^e \wedge F^e) - n_f^2 (F^e \wedge F^e)   ],& \nonumber\\
C^3 \wedge  
[\beta^2 {\tilde \epsilon}] [ 3(F^a \wedge
F^a)- 6 (F^d \wedge F^d) - 4 (F^e \wedge F^e)
- 2 (F^f \wedge F^f)
],& \nonumber\\
\label{nonanomal}
\eeqa

\section{Higgs sector}

\subsection{\em Stability of the configurations and Higgs sector}

We have so far seen the appearance in the R-sector 
of $I_{ab}$ massless fermions
in the D-brane intersections 
transforming under bifundamental representations $N_a, {\bar N}_b$.
 In intersecting 
brane words, besides the 
presence of massless fermions at each intersection, 
we have present of an equal number of
 massive bosons, in the NS-sector, in the same representations 
as the massless fermions \cite{luis1}.
Their mass is of order of the string scale.
However, some of those 
massive bosons may become 
tachyonic \footnote{For consequences
when these set of fields may become massless see \cite{cim}.}, 
especially when their mass, that depends on the 
angles between the branes,
is such that is decreases the world volume of the 
3-cycles involved in the recombination process of joining the two
branes into a single one \cite{senn}.
Denoting the twist vector by $(\vartheta_1,\vartheta_2,
\vartheta_3,0)$, in the NS open string sector the 
lowest lying states are given by \footnote{
we assume $0\leq\vartheta_i\leq 1$ .}
{\small \beqa
\begin{array}{cc}
{\rm \bf State} \quad & \quad {\bf Mass} \\
(-1+\vartheta_1,\vartheta_2,\vartheta_3,0) & \alpha' M^2 =
\frac 12(-\vartheta_1+\vartheta_2+\vartheta_3) \\
(\vartheta_1,-1+\vartheta_2,\vartheta_3,0) & \alpha' M^2 =
\frac 12(\vartheta_1-\vartheta_2+\vartheta_3) \\
(\vartheta_1,\vartheta_2,-1+\vartheta_3,0) & \alpha' M^2 =
\frac 12(\vartheta_1+\vartheta_2-\vartheta_3) \\
(-1+\vartheta_1,-1+\vartheta_2,-1+\vartheta_3,0) & \alpha' M^2
= 1-\frac 12(\vartheta_1+\vartheta_2+\vartheta_3)
\label{tachdsix}
\end{array}
\eeqa}
Exactly at the point, where one of these masses may
become massless we have
preservation of ${\cal N}=1$ SUSY. 
We note that the angles at the four different intersections can be expressed
in terms of the parameters of the tadpole solutions.

$\bullet$ {\em Angle structure and Higgs fields for PS-III classes of models}

The angles at the different intersections can be expressed in terms of the 
tadpole solution parameters. 
We define the angles:
\beqa
\theta_1 \   = \ \frac{1}{\pi}
cot^{-1}\frac{ R_1^{(1)}}{ \epsilon m_b^1 R_2^{(1)}} \ ;\
\theta_2 \  =   \  \frac{1}{\pi} cot^{-1}
\frac{n_a^2 R_1^{(2)}}{3 \epsilon {\tilde \epsilon}\beta_2 R_2^{(2)}} \ ;\
\theta_3 \  = \  \frac{1}{\pi} cot^{-1}\frac{2R_1^{(3)}}{R_2^{(3)}},
 \nonumber \\
{\tilde {\theta_1}} \   = 
\ \frac{1}{\pi} cot^{-1}\frac{ R_1^{(1)}}{\epsilon m_c^1 R_2^{(1)}}\
,\ {\tilde {\theta_2}} \   =
\ \frac{1}{\pi} cot^{-1}\frac{\epsilon
{\tilde \epsilon} n_d^2 R_1^{(1)}}{ 3
 \b_2 R_2^{(1)}}, \
 {\bar {\theta_2}} \   = \
 \frac{1}{\pi} cot^{-1}\frac{\epsilon
 {\tilde \epsilon}n_e^2 R_1^{(1)}}{ 2
 \beta_2 R_2^{(1)}}  \nonumber \\
 \theta_2^{\prime} \   = \
 \frac{1}{\pi} cot^{-1}\frac{\epsilon
 {\tilde \epsilon}n_f^2 R_1^{(1)}}{ 2
 \beta_2 R_2^{(1)}}  
\label{angPSII}
\eeqa
where we consider $\epsilon {\tilde \epsilon} >0$,
$\epsilon m_b^1 >0$, $\epsilon m_c^1 >0$ and 
$R_{i}^{(j)}$, $i={1,2}$ are the compactification radii
for the three $j=1,2,3$ tori, namely
projections 
of the radii 
 onto the cartesian axis $X^{(i)}$ directions when the NS flux B field,
$b^k$, $k=1,2$ is turned on. 

At each of the eight non-trivial intersections 
we have the 
presense of four states $t_i , i=1,\cdots, 4$, that could
become massless, associated
to the states (\ref{tachdsix}).
 Hence we have a total of
thirty two different scalars in the model.
The setup is seen clearly if we look at figure one.
These scalars are generally massive but for some values of
their angles could become tachyonic (or massless).

Also, if we demand that the scalars associated with (\ref{tachdsix}) and 
PS-III models 
may not be tachyonic,
we obtain a total of eighteen 
conditions for the PS-III type models
 with a D6-brane at angles
configuration to be stable.
They are
given in Appendix I.
We don't consider
the scalars from 
the $aa^{\star}$, $dd^{\star}$, $ee^{\star}$,
$ff^{\star}$ intersections. For these sectors
we will require later that they preserve $N=1$ SUSY. 
As a result all scalars in these sectors may become
massive or receive vevs and becoming eventually massive.

%%%%%%%%%% Figure here%%%%%%%%%%%%%%%%%%%%%%%%%%%%%%%%%%%%%%%%%
\begin{figure}
\centering
\epsfxsize=6in
\hspace*{0in}\vspace*{.2in}
\epsffile{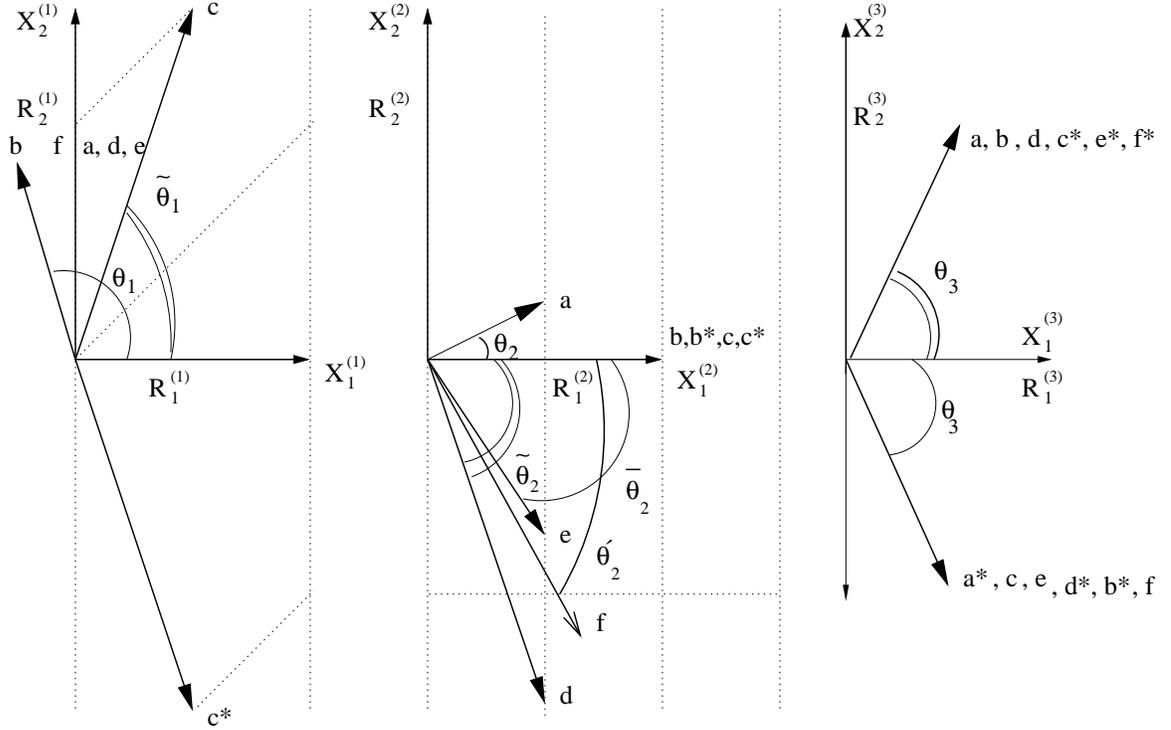}
\caption{\small
Assignment of angles between D6-branes on the
PS-III class of models
based on the initial gauge group $U(4)_C\times {U(2)}_L\times
{U(2)}_R$. The angles between branes are shown on a product of 
$T^2 \times T^2 \times T^2$. We have chosen  $m_b^1, 
m_c^1, n_a^2, n_d^2, n_e^2, n_f^2  >0$,$\epsilon = {\tilde \epsilon}= 1$.
These models
break to low energies to exactly the SM.}
\end{figure}
%%%%%%%%%%%%%%%%%end of figure %%%%%%%%%%%%%%%%%%%%%%%%%%%%%%%%%%%

Lets us now turn our discussion to the Higgs sector of 
PS-III models.
In general there are two different Higgs fields that may
be used to break the PS symmetry.
We remind that they were given in (\ref{useful1}).
The question is if $H_1$, $H_2$ are present in the spectrum of 
PS-III models.
In general, tachyonic scalars stretching between two 
different branes $\tilde a$, 
$\tilde b$, can be used as Higgs scalars as they can become non-tachyonic
by varying the distance between the branes.
Looking at the $I_{a c^{\star}}$ intersection we can confirm that 
the scalar doublets $H^{\pm}$ get localized.
They come from open strings
stretching
between the $U(4)$ $a$-brane and $U(2)_R$ $c^{\star}$-brane.

\begin{table} [htb] \footnotesize
\renewcommand{\arraystretch}{1}
\begin{center}
\begin{tabular}{||c|c||c|c|c|c||}
\hline
\hline
Intersection & PS breaking Higgs & $Q_a$ & $Q_b$ & $Q_c$ & $Q_d$ \\
\hline\hline
$a c^{\star}$ & $H_1 = (4, 1, 2)$  &
$1$ & $0$ & $1$ & $0$ \\
\hline
$a c^{\star}$  & $H_2 = ({\bar 4}, 1, {\bar 2})$  
& $-1$ & $0$ & $-1$ & $0$  \\
\hline
\hline
\end{tabular}
\end{center}
\caption{\small Higgs fields responsible for the breaking of 
$SU(4) \times SU(2)_R$ 
symmetry of the 
$SU(4)_C \times SU(2)_L \times SU(2)_R$ with D6-branes
intersecting at angles. These Higgs are responsible for giving
masses to the right handed
neutrinos in a single family.
\label{Higgs}}
\end{table}

The $H^{\pm}$'s come from the NS sector and
correspond to the states \footnote{a similar set of states was used
in \cite{louis2} to provide the model with electroweak Higgs scalars.}  
{\small \beqa
\begin{array}{cc}
{\rm \bf State} \quad & \quad {\bf Mass^2} \\
(-1+\vartheta_1, \vartheta_2, 0, 0) & \alpha' {\rm (Mass)}^2_{H^{+}} =
  { {Z_3}\over {4\pi ^2}}\ +\ \frac{1}{2}(\vartheta_2 - 
\vartheta_1) \\
(\vartheta_1, -1+ \vartheta_2, 0, 0) & \alpha' {\rm (Mass)}^2_{H^{-}} =
  { {Z_3}\over {4\pi ^2 }}\ +\ \frac{1}{2}(\vartheta_1
  - \vartheta_2) \\
\label{Higgsmasses}
\end{array}
\eeqa}
where $Z_3$ is the distance$^2$ in transverse space along the third torus, 
$\vartheta_1$, $\vartheta_2$ are the (relative)angles 
between the $a$-, $c^{\star}$-branes in the 
first and second complex planes respectively.  
The presence of scalar doublets $H^{\pm}$ can be seen as
coming from the field theory mass matrix

\beq
(H_1^* \ H_2) 
\left(
\bf {M^2}
\right)
\left(
\begin{array}{c}
H_1 \\ H_2^*
\end{array}
\right) + h.c.
\eeq
where
\beqa
{\bf M^2}=M_s^2
\left(
\begin{array}{cc}
Z_3^{(ac^*)}(4 \pi^2)^{-1}&
\frac{1}{2}|\vartheta_1^{(ac^*)}-\vartheta_2^{(ac^*)}|  \\
\frac{1}{2}|\vartheta_1^{(ac^*)}-\vartheta_2^{(ac^*)}| &
Z_3^{(ac^*)}(4 \pi^2)^{-1}\\
\end{array}
\right),
\eeqa
\vspace{1cm}
The fields $H_1$ and $H_2$ are thus defined as
\beq
H^{\pm}={1\over2}(H_1^*\pm H_2) 
\eeq
where their charges are given in table (\ref{Higgs}). 
Hence the effective potential which 
corresponds to the spectrum of the PS symmetry breaking
Higgs scalars is given by
\beqa
V_{Higgs}\ =\ m_H^2 (|H_1|^2+|H_2|^2)\ +\ (m_B^2 H_1H_2\ +\ h.c)
\label{Higgspot}
\eeqa
where
\beqa 
{m_H}^2 \ =\ {{Z_3^{(ac^*)}}\over {4\pi ^2\alpha '}} &;&
m_B^2\ =\ \frac{1}{2\alpha '}|\vartheta_1^{(ac^*)}-
\vartheta_2^{(ac^*)}|
\label{masillas}
\eeqa
The precise values of $m_H^2$, $m_B^2$, for PS-III
classes of
 models are given by
\beqa
 {m_H}^2 \ \stackrel{PS-III}{=}\ {
 {(\xi_a^{\prime}+\xi_c^{\prime})^2}\over {\alpha '}},&&
m_B^2\ \stackrel{PS-III}{=}\
\frac{1}{2\alpha'}|\frac{1}{2} + {\tilde\theta}_1 -
{\theta}_2|\ ,
\label{value100}
\eeqa
where $\xi_a^{\prime}$($\xi_c^{\prime}$) is the distance between the
orientifold plane
and the $a$($c$) branes and ${\tilde \theta}_1$, 
$ {\theta}_2$ were defined in
(\ref{angPSII}). Thus

\beqa
m_B^2 &\stackrel{PS-III}
{=}&
\frac{1}{2}|
m^2_{\chi_R^2} (t_2) +\ m^2_{\chi_R^2}(t_3)
-\ m^2_{{F}_L} (t_1) -\ m^2_{{F}_L}(t_3) |\nonumber\\
&=&
\frac{1}{2}| m^2_{\chi_R^2} (t_2)
+\ m^2_{\chi_R^2 }(t_3)
-m^2_{{\bar F}_R} (t_1) -\ m^2_{{\bar F}_R}(t_3)|
\nonumber\\
\label{kainour}
\eeqa

For PS-III models the number of Higgs present
is equal to the 
the intersection number product between the $a$-, $c^{\star}$- branes
in the
first and second complex planes, 
\beq
n_{H^{\pm}} \stackrel{PS-III}{=}\
I_{ac^{\star}}\ =\ 3.
\label{inter1}
\eeq
A comment is in order.   
For PS-III models the number of PS Higgs is
three.
That means that we have
three
intersections and to each one we have a Higgs particle
which is a linear
combination of the Higgs $H_1$ and $H_2$.

The electroweak symmetry breaking could be delivered through
the bidoublets Higgs present
in the $bc^{\star}$ intersection (seen in table
(\ref{Higgs3}). In principle these 
can be used to give mass ot quarks and leptons.
In the present models their number is
given by
\begin{table} 
%[htb]
% \footnotesize
%\renewcommand{\arraystretch}{2.5}
\begin{center}
\begin{tabular}{||c|c||c|c|c|c||}
\hline
\hline
Intersection & Higgs & $Q_a$ & $Q_b$ & $Q_c$ & $Q_d$ \\
\hline\hline
$bc^{\star}$  & $h_1 = (1, 2, 2)$  &
$0$ & $1$ & $1$ & $0$ \\
\hline
$bc^{\star}$ & $h_2 = (1, {\bar 2}, {\bar 2})$  & $0$ & $-1$ & $-1$ & $0$  \\
\hline
\hline
\end{tabular}
\end{center}                 
\caption{\small Higgs fields present in the intersection $bc^{\star }$
of the
$SU(4)_C \times SU(2)_L \times SU(2)_R$ classes of models
with D6-branes
intersecting at angles. These Higgs give masses to the quarks and
leptons in a single family and could have been
responsible for
electroweak symmetry breaking if their net number was not
zero.
\label{Higgs3}}
\end{table}
the intersection number of the 
$b$, $c^{\star}$ branes in the first tori
\beq
n_{h_1,\  h_2}^{b c^{\star}}\
\stackrel{PS-III}{=}\ |\epsilon(m_c^1 - m_b^1)|  =\
|\beta^2 (2n_a^2 - n_d^2 - n_e^2 -\frac{n_f^2}{2})|
\label{nadou1}
\eeq
The number of the
electroweak
bidoublets in the PS-III models depends on the difference
$|m_b^1-m_c^1|$, taking into account the conditions for N=1 SUSY in some 
sectors, e.g. (\ref{modede1}) in section (7.1),
we get $n_{h^{\pm}} = 0$ and thus $m_b^1 =\ m_c^1$.
However, this is not a problem for electroweak symmetry breaking as 
(see section 8) a different term is used to provide Dirac 
masses to quarks, leptons and neutrinos \footnote{the same conditions hold
for the PS-II models of \cite{kokos3}  
 }. In the present
models is it
important that 
\beq
I_{bc} =\ |m_c^1 + m_b^1| = 2 |m_b^1|
\label{nadou112}
\eeq
may be chosen different from zero.
Thus an alternative set of electroweak Higgs may be
provided from the 
the NS sector where
the lightest scalar states $h^{\pm}$ originate from
open strings stretching between the
$bc$ branes, e.g. named as $h_3$, $h_4$.

\begin{table} 
%[htb]
% \footnotesize
%\renewcommand{\arraystretch}{2.5}
\begin{center}
\begin{tabular}{||c|c||c|c|c|c||}
\hline
\hline
Intersection & Higgs &  $Q_a$ & $Q_b$ & $Q_c$ & $Q_d$ \\
\hline\hline
$bc$  & $h_3 = (1, {\bar 2}, 2)$  &
$0$ & $-1$ & $1$ & $0$ \\
\hline
$bc$ & $h_4 = (1, 2, {\bar 2})$  & $0$ & $1$ & $-1$ & $0$  \\
\hline
\hline
\end{tabular}
\end{center}                 
\caption{\small Higgs fields present in the
intersection $bc$
of the
$SU(4)_C \times SU(2)_L \times SU(2)_R$ classes of models
with D6-branes
intersecting at angles. These Higgs give masses to the
quarks and
leptons in a single family and are
responsible for
electroweak symmetry breaking.
\label{Higgs33}}
\end{table}

{\small \beqa
\begin{array}{cc}
{\rm \bf State} \quad & \quad {\bf Mass^2} \\
(-1+\vartheta_1, 0, \vartheta_3, 0) & \alpha' {\rm (Mass)}^2 =
  \frac{Z_2^{bc}}{4\pi^2}\ +\
  \frac{1}{2}(\vartheta_3 - \vartheta_1)  \\
 (\vartheta_1, 0, -1+\vartheta_3, 0) &  \alpha' {\rm (Mass)}^2 =
  \frac{Z_2^{bc}}{4\pi^2} +\
  \frac{1}{2}(\vartheta_1 -\vartheta_3 )
\label{Higgsacstar}
\end{array}
\eeqa}
where $Z_2^{bc}$ is the relative distance in 
transverse space along the second torus from the orientifold plane, 
$\vartheta_1$, $\vartheta_3$, are the (relative)angle 
between the $b$-, $c$-branes in the 
first and third complex planes.

Hence the presence of scalar doublets $h^{\pm}$
defined as
\beq
h^{\pm}={1\over2}(h_3^*\pm h_4) \   \ .
\eeq
can be seen
as
coming from the field theory mass matrix

\beq
(h_3^* \ h_4) 
\left(
\bf {M^2}
\right)
\left(
\begin{array}{c}
h_3 \\ h_4^*
\end{array}
\right) + h.c.
\eeq
where
\beqa
{\bf M^2}=M_s^2
\left(
\begin{array}{cc}
Z_{23}^{(bc)}(4\pi^2)^{-1}&
\frac{1}{2}|\vartheta_1^{(bc)}-\vartheta_3^{(bc)}|  \\
\frac{1}{2}|\vartheta_1^{(bc)}-\vartheta_3^{(bc)}| &
Z_{23}^{(bc)}(4\pi^2)^{-1}\\
\end{array}
\right),
\eeqa
The effective potential which 
corresponds to the spectrum of electroweak
Higgs $h_3$, $h_4$ may be written as
\beqa
V_{Higgs}^{bc}\ =\ \overline{m}_H^2 (|h_3|^2+|h_4|^2)\ +\ 
(\overline{m}_B^2 h_3 h_4\ +\ h.c)
\label{bcstarpote}
\eeqa
where
\beqa 
\overline{m}_H^2 \ =\ \frac{Z_2^{(bc)}}{4\pi^2\alpha'} \
& ;&
\overline{m}_B^2\ =\ \frac{1}{2\alpha'}|
\vartheta_1^{(bc)} - \vartheta_3^{(bc)}|
\label{bchiggs}
\eeqa
The precise values of
$\overline{m}_H^2$, $\overline{m}_B^2$ are
\beqa 
 {\bar m}_{H}^2 \ \stackrel{PS-III}{=}\ { {({\tilde \chi}_b^{(2)}
 +{\tilde \chi}_{c^{\star}}^{(2)} )^2}\over
{\alpha '}}\ ;\
{\bar m}_{B}^2\ \stackrel{PS-III}{=}\ \frac{1}{2\alpha'}
|\theta_1 - \tilde\theta_1 - 2 \theta_3|
\ ;
\label{value1002}
\eeqa
where $\theta_1$,  $\tilde \theta_1$, $\theta_3$ were defined
in (\ref{angPSII}).
Also ${\tilde \chi}_b, {\tilde \chi}_{c^{\star}}$ are the 
distances of the $b$, $c$ branes from the
orientifold plane in the second tori.
The values of the angles $\vartheta_1$,
$\tilde \vartheta_1$, $\vartheta_2$,
can be expressed
in terms of the scalar masses at the various
intersections
\beqa
\frac{1}{\pi}\theta_1 &=& \frac{1}{2}|
 m^2_{F_L} (t_2) +\ m^2_{F_L}(t_3) -1 |
=  \frac{1}{2}|
m^2_{\chi_L^1} (t_2) +\ m^2_{\chi_L^1}(t_3) + 1|
 \nonumber\\
&=& \frac{1}{2}|
 m^2_{\chi_L^2} (t_2) +\ m^2_{\chi_L^2}(t_3) + 1| 
= \frac{1}{2}|
 m^2_{\chi_L^3} (t_2) +\ m^2_{\chi_L^3}(t_3)+ 1 |
 \label{a1}
\eeqa
\beqa
\frac{1}{\pi}{\tilde \theta}_1 &=&  \frac{1}{2}|
m^2_{{\bar F}_R} (t_2) -\ m^2_{{\bar F}_R }(t_3) - 1|
= \frac{1}{2}|m^2_{\chi_R^1} (t_2)
+\ m^2_{\chi_R^1}(t_3) - 1|
\nonumber\\
&=& \frac{1}{2}|m^2_{\chi_R^2 }
(t_2) +\ m^2_{\chi_R^2  }(t_3) - 1|
= \frac{1}{2}|m^2_{\chi_R^3 }
(t_2) +\ m^2_{\chi_R^3 }(t_3) - 1|
\label{a2}
\eeqa
\beqa
\frac{1}{\pi}\theta_2 &=& \frac{1}{2}|
 m^2_{F_L} (t_1) +\ m^2_{F_L}(t_3) | 
= \frac{1}{2}| m^2_{{\bar F}_R} (t_1) +\ m^2_{{\bar F}_R}(t_3) |
\eeqa
\beqa
\frac{1}{\pi}\theta_3 &=& \frac{1}{4}|
 m^2_{F_L} (t_1) +\ m^2_{F_L}(t_2) | = \frac{1}{4}|
 m^2_{{\bar F}_R} (t_1) +\ m^2_{{\bar F}_R}(t_2) | \nonumber\\
&=& \frac{1}{4}|
 m^2_{\chi_L^j} (t_1) +\ m^2_{\chi_L^j}(t_2) |
 = \frac{1}{4}|
 m^2_{\chi_R^j} (t_1) +\ m^2_{ \chi_R^j}(t_2)|, \ j=1,2,3 \nonumber\\
 \label{a3}
\eeqa

\section{Singlet scalar generation - $N=1$ SUSY
on Intersections}

In this section, we will
present a gauge singlet generation mechanism
by demanding that
certain open string sectors
respect $N=1$ supersymmetry.
Similar considerations
were useful on the other GUT classes of PS-like
models with only the SM at low energy
\cite{kokos1, kokos2, kokos3} and also
in the models from a SM-like configuration at $M_s$
with the number of stacks being only five and six
respectively \cite{kokos, kokoss}.
The singlet scalars
will be necessary for giving
masses to the $U(1)$'s which they don't couple to
the RR fields.
Also they will be used to realize
a Majorana mass term
for the right handed
neutrinos.
We note that the spectrum
of PS-III classes of models described in
table (\ref{spectrum8}) is massless at this point.
Supersymmetry will create singlet scalars
which receive vevs and generate masses for the 
otherwise massless
fermions $\chi_L^1$, $\chi_L^2$, $\chi_L^3$,
$\chi_R^1$, $\chi_R^2$, $\chi_R^3$,
$\omega_L$, $y_R$, $s_R^1$, $s_R^2$, $s_R^3$.

\subsection{\em PS-III models with N=1 SUSY}

In this part we will show that model dependent
conditions,
obtained by demanding that the extra U(1)'s do not have
non-zero couplings to the RR fields, are necessary
conditions in order to have scalar singlet generation
that could effectively break the extra U(1)'s.
These conditions will be alternatively obtained by demanding that certain
string sectors respect $N=1$ supersymmetry.

In general, for $N=1$ supersymmetry to be
preserved at some
intersection
between two branes {\em L}, {\em M}, we need to
satisfy
$\pm \vartheta_{ab}^1 \pm \vartheta_{ab}^2 \pm
\vartheta_{ab}^3$ for
some choice of signs, where
$\vartheta_{\alpha \beta}^i$, $i=1,2,3$
are the relative angles of the
branes {\em L}, {\em M} across the three 2-tori.
The latter rule will be our main tool in getting
N=1 SUSY on intersections.

\begin{itemize}

\item {\em The $ac$ sector respects ${\cal N} =1$
supersymmetry.}

The condition for $N=1$ SUSY on the $ac$-sector
is \footnote{We have chosen $m_c^1 < 0$.}:
\beq
\pm (\frac{\pi}{2} +\ {\tilde \vartheta}_1) \
\pm  {\vartheta}_2 \ \pm  2\vartheta_3 \ = 0,
\label{condo1}
\eeq       
This condition can be solved by choosing :
\beq
ac \rightarrow (\frac{\pi}{2} + \ {\tilde \vartheta}_1) \
+ \vartheta_2 \
- 2\vartheta_3 \ = 0,
\label{condo10}
\eeq       
and thus may be solved by the choice \footnote{We have set
$U^{(i)}= \frac{R_2^{(i)}}{R_1^{(i)}}$, $i=1, 2,3$}
\beq
-{\tilde \vartheta}_1 \ =
\vartheta_2 \ = \vartheta_3 \ = \frac{\pi}{4},
\label{solver1}
\eeq
effectively giving us
\beq
-  \frac{1}{\epsilon m_c^1 \ U^{(1)}} = \
\frac{(\epsilon {\tilde \epsilon}) n_a^2}{3  \b_2 U^{(2)}}
= \
  \frac{2 {\tilde \epsilon}}{U^{(3)}} = \ \frac{\pi}{4}.
\label{condo3}
\eeq
By imposing $N=1$ SUSY on an intersection ac the 
massless scalar
superpartner of ${\bar F}_R$ appears, the ${\bar F}_R^B$.
Note that in (\ref{condo3}) the imposition of N=1 SUSY
connects the complex structure moduli $U^i$ in the different tori
and thus reduces the moduli degeneracy of the theory.

\end{itemize}

\begin{itemize}

\item {\em The $dd^{\star}$ sector preserves
${\cal N} = 1$ supersymmetry}

As we noted in the appendix the presence of N=1
supersymmetry in the sectors $dd^{\star}$,
$ee^{\star}$ is equivalent to the absence of tachyons in
those sectors.

The general form of the ${\cal N} = 1$
supersymmetry condition on this sector
is
\beq
\pm \pi \pm 2 {\tilde \vartheta}_2 \ \pm  2
{\vartheta}_3 \
\ = 0,
\label{condo1oup1}
\eeq       
which may be solved by the choice
\beq
- \pi \ + 2 {\tilde \vartheta}_2 \ +  2
{\vartheta}_3 \
\ = 0,
\label{esosi1}
\eeq       
Hence
\beq
{\tilde \vartheta}_2 \ = \vartheta_3 \ = \frac{\pi}{4},
\label{solv1sol23}
\eeq
that is
\beq
\frac{\epsilon {\tilde \epsilon}n_d^2}{3 \beta_2} U^{(2)} = \
  \frac{2}{{\tilde \epsilon}}U^{(3)} = \ \frac{\pi}{4}.
\label{laop13}
\eeq

From (\ref{condo3})  and  (\ref{laop13}) we deduce that
\beq
n_a^2 = n_d^2
\eeq

\end{itemize}

Due to the presence of N=1 SUSY on this sector we have localized
the superpartner of the $s_R^1$, the $s_B^1$.

\begin{itemize}

\item {\em The $ee^{\star}$ sector preserves
${\cal N} = 1$ supersymmetry}

The general form of the ${\cal N} = 1$
supersymmetry condition on this sector
is
\beq
\pm \pi \pm\ 2 {\bar \vartheta}_2 \ \pm  2
{\vartheta}_3 \
\ = 0,
\label{oppprae1}
\eeq       
which we may recast in the form
\beq
- \pi +\ 2 {\bar \vartheta}_2 \ + 2
{\vartheta}_3 \
\ = 0,
\label{allo2}
\eeq       
be solved by the choice
\beq
{\bar  \vartheta}_2 \ = \vartheta_3 \ = \frac{\pi}{4},
\label{condo23}
\eeq
that is
\beq
\frac{ (\epsilon {\tilde \epsilon}) n_e^2}{2 \beta_2 U^{(2)}} = \
  \frac{2}{{\tilde \epsilon}}U^{(3)} = \ \frac{\pi}{4}.
\label{laopzop13}
\eeq

\end{itemize}
Due to the presence of N=1 SUSY on this sector we have localized
the superpartner of the $s_R^2$, the $s_B^2$.

\begin{itemize}

\item {\em The $ff^{\star}$ sector preserves
${\cal N} = 1$ supersymmetry}

The general form of the ${\cal N} = 1$
supersymmetry condition on this sector
is
\beq
\pm \pi \pm\ 2 {\vartheta}_2^{\prime} \ \pm  2
{\vartheta}_3 \
\ = 0,
\label{oppprae11}
\eeq       
which we may recast in the form
\beq
- \pi +\ 2 \theta_2^{\prime} \ + 2
{\vartheta}_3 \
\ = 0,
\label{allo21}
\eeq       
be solved by the choice
\beq
\theta_2^{\prime} \ = \vartheta_3 \ = \frac{\pi}{4},
\label{condo231}
\eeq
that is
\beq
\frac{ (\epsilon {\tilde \epsilon}) n_f^2}{2 \beta_2 U^{(2)}} = \
  \frac{2}{{\tilde \epsilon}}U^{(3)} = \ \frac{\pi}{4}.
\label{laopzop14}
\eeq

From (\ref{laop13}), (\ref{condo3}),  (\ref{laopzop13}), (\ref{laopzop14}),
 we derive the conditions
\beq 
2 n_a^2 = 3 n_e^2 = 3 n_f^2 = 2 n_d^2 \ . 
\label{modede1}
\eeq
These conditions solve exactly the conditions
(\ref{constr11}).

\end{itemize}
Due to the presence of N=1 SUSY on this sector we have localized
the superpartner of the $s_R^3$, the $s_B^3$.

Thus the presence
of N=1 supersymmetry in $dd^{\star}$, $ee^{\star}$,
$ff^{\star}$ sectors guarantees the presence of gauge singlets as 
scalar
superpartners of the $s_R^1$, $s_R^2$, $s_R^3$ fermions, e.g. 
$s_B^1$, $s_B^2$, $s_B^3$ that may may
receive vevs of undetermined order.

Also what is is evident by looking at conditions
(\ref{constr11}), (\ref{modede1}) 
is that the
conditions of orthogonality for the extra $U(1)$'s
to survive massless the generalized Green-Schwarz
mechanism is equivalent to the conditions for N=1
supersymmetry in the leptonic sectors $dd^{\star}$,
$ee^{\star}$.  
The latter condition is equivalent to the
absence of tachyons in the sectors $dd^{\star}$,
$ee^{\star}$.

\subsection{\em Gauge singlet generation from the extra
U(1) branes}

In this section, we will present an alternative
mechanism for generating singlet scalars.
We had already seen that in leptonic sectors involving
U(1) branes, e.g. $dd^{\star}$, $ee^{\star}$, brane
imposing
N=1 SUSY creates singlet scalars. This is reflected
in the fact that in U(1) $j$-branes, sectors in the form
$jj^{\star}$ had localized in their intersection
gauge singlet fermions. Thus imposing N=1 SUSY on those
sectors help us to get rid of these massless fermions,
by making them
massive through their couplings to their superpartner
gauge singlet scalars.

What we will become clear in this sector is that the
presence of supersymmetry in particular sectors involving
the extra branes creates singlet scalars that 
provide the couplings that make massive
some non-SM fermions.

In order to show the creation of gauge singlets
from sectors involving extra branes we will make our
points
by using only one of the extra $N_h$ $U(1)$ branes, e.g. the
$N_{h_1}$ one. The following
discussion can be identically repeated for the other
extra branes.

Thus due to the non-zero intersection numbers
of the
$N_{h_1}$ U(1) brane with $a$,$d$ branes the following
sectors are
present : $ah$, $ah^{\star}$, $dh$, $dh^{\star}$.

\begin{itemize}

\item $ah$-{\em sector}

Because $I_{ah} = -\frac{3}{\beta_1}$
we have present $|I_{ah}|$ massless fermions
$\kappa_1^f$ in the representations
\beq
\kappa_1^f \rightarrow \ ({\bar 4}, 1, 1)_{(-1, 0, 0, 0, 0, 0; 1)}
\label{ah}
\eeq
where the subscript last entry denotes the U(1) charge
of the of the U(1) extra brane \footnote{We don't exhibit
the beyond the seventh entry of the rest of the extra
branes as for the present discussion are identically zero.}.

\item $ah^{\star}$-{\em sector}

Because $I_{ah^{\star}}= -\frac{3}{\beta_1} $,
there are
$|I_{ah^{\star}}|$
fermions $\kappa_2^f$ localized in the
$ah^{\star}$ intersection and
appearing as 
\beq
\kappa_2^f \rightarrow \ ({\bar  4}, 1, 1 )_{(-1,
0, 0, 0, 0, 0; -1)}
\l{dh2}
\eeq

\item $dh$-{\em sector}

Because $I_{dh} = -\frac{6}{\beta_1}$, there are
present
$|I_{dh}|$ fermions $\kappa_3^f$ transforming in the representations
\beq
\kappa_3^f \rightarrow \ (1, 1, 1)_{(0, 0, 0, -1, 0, 0; -1)}
\label{reps1}
\eeq

We further require that this sector respects
$N=1$ supersymmetry.
In this case we have also present the massless scalar
fields $\kappa_3^B$,
\beq
\kappa_3^B \rightarrow \ (1, 1, 1 )_{(0, 0, 0, -1, 0, 0; -1)},
\l{aradh5}
\eeq
The latter scalars receive a vev which we assume to be
of order of the string scale.

 The condition for
$N=1$ supersymmetry in this sector is exactly
\beq
-\frac{\pi}{2} +\ {\tilde \vartheta}_2
+\ \vartheta_3 = 0
\label{exactamente1}
\eeq
which is satisfied when ${\tilde \vartheta}_2$,
${\vartheta}_3$ take the value $\pi/4$ in consistency
with (\ref{solv1sol23}) and subsequently (\ref{allo2}).

\item $dh^{\star}$-{\em sector}

Because $I_{dh^{\star}} =  -\frac{6}{\beta_1} \neq 0$,
there are present
$|I_{ah^{\star}}|$
fermions $\kappa_4^f$ in the representations
\beq
\kappa_4^f \rightarrow \ ( 1, 1, 1 )_{(0, 0, 0, -1, 0, 0; -1)}
\l{dh4}
\eeq

The condition that this sector respects N=1 SUSY is
equivalent to the one is the $dh$-sector.

\item $eh$-{\em sector}

Because $I_{eh} = -\frac{4}{\beta_1}$, there are
present
$|I_{eh}|$ fermions $\kappa_5^f$ transforming in the
representations
\beq
\kappa_5^f \rightarrow \ (1, 1, 1)_{(0, 0, 0, 0, -1, 0; -1)}
\label{erepsec1}
\eeq

Also we require that this sector
preserves N=1 SUSY.
Because of the presence of N=1 SUSY
there is evident the presence
of $|I_{eh}|$ bosons $\kappa_5^B$ transforming in the
representations
\beq
\kappa_5^B \rightarrow \ (1, 1, 1)_{(0, 0, 0, 0, -1, 0; 1)}
\label{boso1}
\eeq
The condition for N=1 SUSY is
\beq
\pm \frac{\pi}{2} \pm {\bar \vartheta}_2 \pm \vartheta_3 =\ 0
\label{halfcond1}
\eeq
which is exactly `half' of the supersymmetry condition
(\ref{allo2}). When it is rearranged into the form
\beq
\frac{\pi}{2} + {\bar \vartheta}_2 - \vartheta_3 = \ 0, 
\label{reaa2}
\eeq
it is solved by the choice (\ref{condo23}).

\item $eh^{\star}$-{\em sector}
In this sector,  $I_{eh^{\star}} = -\frac{4}{\beta_1}$.
 Thus there are
present
$|I_{eh^{\star}}|$ fermions $\kappa_6^f$ transforming in
the
representations
\beq
\kappa_6^f \rightarrow \ (1, 1, 1)_{(0, 0, 0, 0, -1, 0; -1)}
\label{erepsec112}
\eeq
The condition for N=1 SUSY to be preserved by this section
is exactly (\ref{halfcond1}). Thus we have present
$|I_{eh^{\star}}|$ bosons $\kappa_6^B$ transforming
in the 
representations
\beq
\kappa_6^B \rightarrow \ (1, 1, 1)_{(0, 0, 0, 0, -1, 0; -1)}
\label{boso2}
\eeq

\item $fh$-{\em sector}

Because $I_{fh} = -\frac{2}{\beta_1}$, there are
present
$|I_{fh}|$ fermions $\kappa_7^f$ transforming in the
representations
\beq
\kappa_7^f \rightarrow \ (1, 1, 1)_{(0, 0, 0, 0, 0, -1; 1)}
\label{erepsec13}
\eeq

Also we require that this sector
preserves N=1 SUSY.
Because of N=1 SUSY 
there are
present
$|I_{fh}|$ bosons $\kappa_7^B$ transforming in the
representations
\beq
\kappa_7^B \rightarrow \ (1, 1, 1)_{(0, 0, 0, 0, 0, -1; 1)}
\label{bosoon1}
\eeq
The condition for N=1 SUSY is
\beq
\pm \frac{\pi}{2} \pm {\vartheta}_2^{\prime} \pm \vartheta_3 =\ 0
\label{halfcond11}
\eeq
which is `half' of the supersymmetry condition
(\ref{allo21}). When it is rearranged into the form
\beq
-\ \frac{\pi}{2} +\ {\vartheta}_2^{\prime} +\ \vartheta_3 = \ 0, 
\label{reaa23}
\eeq
it is solved by the choice (\ref{condo231}).
\newline

\item $fh^{\star}$-{\em sector}
In this sector,  $I_{fh^{\star}} = -\frac{2}{\beta_1}$.
 Thus there are
present
$|I_{fh^{\star}}|$ fermions $\kappa_8^f$ transforming in
the
representations
\beq
\kappa_8^f \rightarrow \ (1, 1, 1)_{(0, 0, 0, 0, 0, -1; -1)}
\label{erepsec115}
\eeq
The condition for N=1 SUSY to be preserved by this section
is exactly (\ref{reaa23}). Thus we have present
$|I_{fh^{\star}}|$ bosons $\kappa_8^B$ transforming
in the 
representations
\beq
\kappa_8^B \rightarrow \ (1, 1, 1)_{(0, 0, 0, 0, -1; -1)}
\label{boso21}
\eeq

\end{itemize}

What we have found \footnote{Similar conclusions
cold be reached for the 5-stack GUT models of \cite{}.}  
 we have found that the conditions
(\ref{constr11}) 
derived as the model dependent conditions
of the U(1)'s that survive the generalized Green-Schwarz
mechanism, are equivalent :

\begin{itemize}

\item to have the leptonic branes, d, e, f, preserve N=1
SUSY on the sectors $dd^{\star}$, $ee^{\star}$,
$ff^{\star}$.
\item to have the sectors made of a mixture of the
extra and leptonic branes preserve N=1 SUSY. The presence of 
these conditions is independent from the number of
extra U(1) branes present.

\end{itemize}

We will now show that all fermions, appearing from
the non-zero intersections of the extra
brane $U(N_{h_1})$
with the branes $a$, $d$,  $e$, receive string scale mass and disappear
from the low energy spectrum (see also a related discussion
in the concluding section).

\begin{itemize}

\item The mass term for the $\kappa_1^f$ fermion reads:
\beqa
(4, 1, 1 )_{(1, 0, 0, 0, 0, 0; -1)} \
(4, 1, 1 )_{(1, 0, 0, 0, 0, 0; -1)}  \
\langle({\bar 4}, 1, 2)_{(-1, 0, 1, 0, 0, 0;0)} \rangle \nonumber\\
\times \langle ({\bar 4}, 1, {\bar 2})_{(-1, 0, -1, 0, 0, 0; 0)} \rangle
\langle (1, 1, 1)_{(0, 0, 0, 0, 0, 1; 1)}\rangle \
\langle (1, 1, 1)_{(0, 0, 0, 0, 0, -1; 1)}\rangle
\label{eksiso1}
\eeqa
or
\beq
{\bar \kappa}_1^f \ {\bar \kappa}_1^f \
\langle H_2 \rangle \
\langle {\bar F}_R^H \rangle \
\langle {\bar \kappa}^8_B \rangle \
\langle {\kappa}^7_B \rangle
\sim    {\bar \kappa}_1^f \ {\bar \kappa}_1^f \    M_s
\eeq

\item The mass term for the $\kappa_2^f$ fermion reads:

\beqa
(4, 1, 1 )_{(1, 0, 0, 0, 0, 0; 1)} \
(4, 1, 1 )_{(1, 0, 0, 0, 0, 0; 1)}  \
\langle({\bar 4}, 1, 2)_{(-1, 0, 1, 0, 0, 0; 0)} \rangle \nonumber\\
\times \langle ({\bar 4}, 1, {\bar 2})_{(-1, 0, -1, 0, 0, 0; 0)} \rangle
\langle (1, 1, 1)_{(0, 0, 0, 0, 0, -1; -1)}\rangle \
\langle (1, 1, 1)_{(0, 0, 0, 0, 0, 1; -1)}\rangle
\label{eksiso2}
\eeqa
or
\beq
{\bar \kappa}_2^f \ {\bar \kappa}_2^f \
\langle H_2 \rangle \
\langle {\bar F}_R^H \rangle \
\langle {\bar \kappa}^7_B \rangle \
\langle {\kappa}^8_B \rangle
\sim    {\bar \kappa}_2^f \ {\bar \kappa}_2^f \    M_s
\eeq

\item The mass term for the $\kappa_3^f$ fermion reads:

\beqa
(1, 1, 1 )_{(0, 0, 0, 0, 1, 0; 1)} \
(1, 1, 1 )_{(0, 0, 0, 0, 1, 0;  1)} \
\langle (1, 1, 1)_{(0, 0, 0, 0, -1, 0; -1)} \rangle  \nonumber\\
\times \langle (1, 1, 1)_{(0, 0, 0, 0, -1, 0; -1)} \rangle 
\label{eksiso4}
\eeqa
or
\beq
{\bar \kappa}_3^f \ {\bar \kappa}_3^f \
\langle{ \kappa}_3^B \rangle \
\langle {\kappa}_3^B \rangle  \sim \ M_s \ {\bar \kappa}_3^f \ 
{\bar \kappa}_3^f
 \eeq

\item The mass term for the $\kappa_4^f$ fermion reads:

\beqa
(1, 1, 1 )_{(0, 0, 0, -1, 0; -1)} \
(1, 1, 1 )_{(0, 0, 0, -1, 0;  -1)} \
\langle (1, 1, 1)_{(0, 0, 0, 1, 0; 1)} \rangle  \nonumber\\
\times \langle (1, 1, 1)_{(0, 0, 0, 1, 0; 1)} \rangle
\label{eksiso44}
\eeqa
or
\beq
{\bar \kappa}_4^f \ {\bar \kappa}_4^f \
\langle{ \kappa}_4^B \rangle \
\langle {\kappa}_4^B \rangle  \sim \ M_s \ {\bar \kappa}_4^f \ 
{\bar \kappa}_4^f
 \eeq

\item The mass term for the $\kappa_5^f$ fermion reads:

\beqa
(1, 1, 1 )_{(0, 0, 0, 0, 1, 0; -1)} \
(1, 1, 1 )_{(0, 0, 0, 0, 1, 0; -1)} \
\langle (1, 1, 1)_{(0, 0, 0, 0, -1, 0; 1)} \rangle  \nonumber\\
\times \langle (1, 1, 1)_{(0, 0, 0, 0, -1, 0; 1)} \rangle
\label{eksiso5}
\eeqa
or
\beq
{\bar \kappa}_5^f \ {\bar \kappa}_5^f \
\langle{ \kappa}_5^B \rangle \
\langle {\kappa}_5^B \rangle  \sim \ M_s \ {\bar \kappa}_5^f \ 
{\bar \kappa}_5^f
 \eeq

\item The mass term for the $\kappa_6^f$ fermion reads:

\beqa
(1, 1, 1 )_{(0, 0, 0, 0, 1, 0; 1)} \
(1, 1, 1 )_{(0, 0, 0, 0, 1, 0; 1)} \
\langle (1, 1, 1)_{(0, 0, 0, -1, 0, 0; -1)} \rangle  \nonumber\\
\times \langle (1, 1, 1)_{(0, 0, 0, -1, 0, 0; -1)} \rangle
\label{eksiso6}
\eeqa
or
\beq
{\bar \kappa}_6^f \ {\bar \kappa}_6^f \
\langle{ \kappa}_6^B \rangle \
\langle {\kappa}_6^B \rangle  \sim \ M_s \ {\bar \kappa}_6^f \ 
{\bar \kappa}_6^f
 \eeq

\item The mass term for the $\kappa_7^f$ fermion reads:

\beqa
(1, 1, 1 )_{(0, 0, 0, 0, 1, 0; 1)} \
(1, 1, 1 )_{(0, 0, 0, 0, 1, 0; 1)} \
\langle (1, 1, 1)_{(0, 0, 0, -1, 0, 0; -1)} \rangle  \nonumber\\
\times \langle (1, 1, 1)_{(0, 0, 0, -1, 0, 0; -1)} \rangle
\label{eksiso61}
\eeqa
or
\beq
{\bar \kappa}_7^f \ {\bar \kappa}_7^f \
\langle{ \kappa}_7^B \rangle \
\langle {\kappa}_7^B \rangle  \sim \ M_s \ {\bar \kappa}_7^f \ 
{\bar \kappa}_7^f
 \eeq

\item The mass term for the $\kappa_8^f$ fermion reads:

\beqa
(1, 1, 1 )_{(0, 0, 0, 0, 1, 0; 1)} \
1, 1, 1 )_{(0, 0, 0, 0, 1, 0; 1)} \
\langle (1, 1, 1)_{(0, 0, 0, -1, 0, 0; -1)} \rangle  \nonumber\\
\times \langle (1, 1, 1)_{(0, 0, 0, -1, 0, 0; -1)} \rangle
\label{eksiso62}
\eeqa
or
\beq
{\bar \kappa}_8^f \ {\bar \kappa}_8^f \
\langle{ \kappa}_8^B \rangle \
\langle {\kappa}_8^B \rangle  \sim \ M_s \ {\bar \kappa}_8^f \ 
{\bar \kappa}_8^f
 \eeq

\end{itemize}

\subsection{\em Breaking the anomaly free massless U(1)'s}

After breaking the PS gauge symmetry at $M_{GUT}$ the
initial gauge symmetry $SU(4)_c \times SU(2)_L
\times SU(2)_R \times U(1)_a \times U(1)_b
\times U(1)_c \times U(1)_d \times U(1)_e \times U(1)_f$
breaks to the SM gauge group
$SU(3) \times SU(2) \times U(1)_Y$ augmented by the extra
anomaly free U(1)'s  
$Q^{(4)}$, $Q^{(5)}$, $Q^{(6)}$, $Q^{(7)}$, $Q^{(8)}$.
The last two are the hidden U(1)'s needed to satisfy the RR
tadpole cancellation conditions. The extra U(1)'s may be
broken if appropriate singlets are available.
The latter may be created by appropriate choosing
the angle parameters between the D6-branes when
demanding N=1 supersymmetry to be preserved in particular
sectors. In this way,
$U(1)^{(4)}$ may be broken if
$s_B^1$ gets a vev,
$U(1)^{(5)}$ may be broken if
$s_B^2$ gets a vev,
$U(1)^{(6)}$ may be broken if
$s_B^3$ gets a vev.
Also $U(1)^{(7)}$ and $U(1)^{(8)}$ may be broken if one
of the
$\kappa_3^B$, $\kappa_4^B$, $\kappa_5^B$, $\kappa_6^B$,
$\kappa_7^B$, $\kappa_8^B$ gets a vev. 
Thus the  U(1)'s surviving massless the Green-Schwarz mechanism 
may be broken easily by using singlets localized on 
intersections of the extra branes and leptonic branes
as well
from $jj^{\star}$ sectors.

We note that up to this point the only issue remaining
is how we can give non-zero masses to all exotic
fermions of
table (\ref{spectrum8}) beyond those that accommodate
the quarks and leptons of the SM.

\section{Yukawa couplings, neutrino and lepton masses}

In this section, we discuss the issue of 
neutrino masses in the
$SU(4) \times S(2)_L \times SU(2)_R$ classes of PS-III
GUTS. Also, we discuss in which way the additional
fermions of table (1), receive a mass and disappear
from the SM spectrum at the $M_Z$ scale.

\subsection{\em Yukawa couplings and Neutrino masses}

Proton decay is the most important problem 
of grand unified theories. In the usual versions of
left-right
symmetric PS models this problem  
is avoided as B-L is a gauged symmetry but the
problem is embedded in
baryon number violating operators of sixth order,
contributing to proton
decay. In the PS-III models there is no-proton decay
as baryon
number is a gauged symmetry, the
corresponding gauge boson becomes massive through its
couplings to RR fields, and thus
 survives as a global symmetry to low energies.
That is a plausible explanation for the origin of
proton stability in general brane-world scenarios.
Baryon B and lepton L numbers are related by $Q_a = 3 B +
L$ and are given by
\beqa
B = \frac{Q_a + Q_{B-L}}{4}.&  
\label{ba1}
\eeqa

For intersecting brane worlds the usual tree level
SM fermion mass generating trilinear Yukawa couplings
between the fermion states
$F_L^i$, ${\bar F}_R^j$ and the Higgs fields $H^k$ depends
on the
stretching of the worldsheet area between the three
D6-branes which cross
at those intersections. In the present Pati-Salam GUTS
the trilinear Yukawa
is
\beq
Y^{ijk} F_L^i {\bar F}_R^j h^k
\label{roof1}
\eeq
For a six dimensional torus in the
leading order \cite{luis1} we have,
\beq
Y^{ijk}=e^{- {\tilde A}_{ijk}},
\label{yuk1}
\eeq
where ${\tilde A}_{ijk}$ is the worldsheet
area \footnote{We note that it is a general property of
string
theories for their Yukawa couplings to depend
exponentially
on the worldsheet area.}
connecting the three vertices.
The areas of each of the $T^2$ tori
taking part in
this interaction are typically of order one in string
units.
In \cite{kokos1} e.g. we have
assumed
that the areas of the second and third tori are close to
zero.
Thus in this case, the area of the full
Yukawa coupling (\ref{yuk1})
may be given in the form 
\beq
Y^{ijk}= e^{-\frac{R_1 R_2}{a^{\prime}} A_{1}},
\label{yuk12}
\eeq
where $R_1$, $R_2$ the radii and 
$ A_{ijk}$ the area of the two dimensional tori in the
first complex plane.
Here we exhibit the leading worldsheet correction coming
from the first tori \footnote{ The same hypothesis
holds for any PS GUT model
constructed so far, as a deformation of the quark and lepton
intersection numbers, e.g. the PS-A, PS-I, PS-II classes of
\cite{kokos1, kokos2, kokos3} respectively.}.

For the present class of GUTS we have seen that
the electroweak
bidoublets (\ref{roof1}) are absent at tree level.
However there is another coupling, which
is non-renormalizable and of the same order :      
\beq
F_L \ {\bar F}_R \
\langle h_3 \rangle \
\langle F_R^H \rangle \langle H_2 \rangle \ \sim \
\upsilon  \
F_L \ {\bar F}_R
\label{blue1}
\eeq

For a dimension five interaction term, like those
involved in the
 Majorana mass term for the right handed neutrinos
the interaction term is in the form
\beq
Y^{lmni}= e^{- {\tilde A}_{lmni}},
\label{yuk14}
\eeq
where ${\tilde A}_{lmni}$ the worldsheet area
connecting the four interaction
vertices. Assuming that the areas
of the second and third tetragonal are
close to zero the four term coupling may be
approximated as
\beq
Y^{ijk}= e^{-\frac{R_1 R_2}{a^{\prime}} A_{2}},
\label{yuk15}
\eeq
where the area of the $A_{2}$ may be of order one in
string units.
\newline
A Majorana mass term for right neutrinos
appears only once we impose $N=1$ SUSY on
an intersection. As a result the
massless scalar
superpartners
of the ${\bar F}_R$ fermions, the ${\bar F}_R^H$'s
appears, allowing the dimension five
Majorana mass term for $\nu_R$,
$F_R F_R {\bar F}_R^H {\bar F}_R^H$.
Hence the full Yukawa interaction for the fermionic
spectrum is 
\beq
\lambda_1 F_L \ {\bar F}_R \ \langle h_3 \rangle \
\langle F_R^B \rangle \langle H_2 \rangle \
+\ \lambda_2 \frac{F_R {F}_R \langle {\bar F}_R^H \rangle
\langle {\bar F}_R^H \rangle}{M_s},
\label{era1} 
\eeq
where
\beqa
\lambda_1 \equiv e^{-\frac{R_1 R_2 A_1}{\alpha^{\prime}}},&
\lambda_2 \equiv e^{-\frac{R_1 R_2 A_2}{\alpha^{\prime}}}.
\label{aswq123}
\eeqa
and the Majorana coupling involves the massless
scalar   
superpartners ${\bar F}_R^H$. 
 The ${\bar F}_R^H$ has a neutral direction that
 receives the vev $<H>$.
There is no restriction on its vev
from first
principles and its vev can be anywhere between
the scale of electroweak symmetry breaking and $M_s$.
\newline
The Yukawa term
\beqa
F_L \ {\bar F}_R \
\langle h_3 \rangle \
\langle F_R^B \rangle \ \langle H_2 \rangle \ \sim \
\upsilon  \
F_L \ {\bar F}_R
\label{yukbre12}
\eeqa
is responsible for the electroweak 
symmetry breaking. This term generates Dirac
masses to up quarks and
neutrinos.
Thus we get
\beq
\lambda_1 F_L \ {\bar F}_R \
\langle h_3 \rangle \ \
\langle F_R^B \rangle \ \langle H_2 \rangle \
  \rightarrow (\lambda_1 \  \upsilon)
(u_i u_j^c + \nu
_i N_j^c) + (\lambda_1 \  {\tilde \upsilon})
\cdot (d_i d_j^c + e_i e_j^c),  
\label{era2}
\eeq
where we have assumed that 
\beq
\langle h_3 \rangle \
\langle F_R^B \ \rangle \langle H_2 \rangle \
= \left(
\begin{array}{cc}
\upsilon  & 0 \\
0 & {\bar \upsilon}
\label{era41}
\end{array}
\right)
\label{finalhiggs}
\eeq
These mass relations may be retained at tree level only,
since as the models
are non-supersymmetric, they will
receive higher order corrections.  
Interestingly from (\ref{finalhiggs})
we derive the GUT relation
\cite{ellis}
\beq
m_d =\ m_e \ .
\label{gutscale}
\eeq
as well the unnatural
 \beq
m_u =\ m_{N^c \nu} \ .
\label{gutscale1}
\eeq 
In the case of neutrino masses, 
the unnatural (\ref{gutscale1}), associated
to the $\nu - N^c$
mixing,
is modified due to the presence of the Majorana term
(\ref{era1})
leading to the see-saw neutrino mass
matrix, of an extended Frogatt- Nielsen type mixing
light with heavy states, 
\beqa
\left(
\begin{array}{cc}
\nu&N^c 
\end{array}
\right)\times
 \left(
\begin{array}{cc}
0  & m \\
m & M
\label{era4}
\end{array}
 \right)
\times
\left(
\begin{array}{c}
\nu\\
N^c 
\label{era5}
\end{array}
\right),
\label{er1245}
\eeqa
where
\beq
m= \lambda_1  \upsilon.
\label{eigen1}
\eeq
After diagonalization
the neutrino mass matrix gives us two eigenvalues,
the heavy eigenvalue
\beq
m_{heavy} \approx M =\ \lambda_2 \frac{<H>^2 }{M_s},
\label{neu2}
\eeq
corresponding to the right handed neutrino and 
the light eigenvalue
\beq
m_{light} \approx \frac{m^2}{M} =\ \frac{\lambda_1^2}{\lambda_2  }
\times\frac{\upsilon^2 \ M_s  } { <H>^2} 
\label{neu1}
\eeq
corresponding
to the left handed neutrino 
Values of the parameters giving us values
for neutrino
masses between 0.1-10 eV, consistent
with the observed neutrino mixing in neutrino
oscillation measurements, 
have already been considered in \cite{kokos1}.
The analysis is identical and it will not repeated here.
We note that
the hierarchy of neutrino masses
has been investigated by examining several
 scenarios associated with a light
$\nu_L$ mass including the cases
$ \langle H \rangle = |M_s|$,
$ \langle H \rangle < | M_s |$.
In both cases the hierarchy of neutrino masses
is easily obtained.

\subsection{\em Exotic fermion couplings}

Up to this point we have shown that all the additional U(1)'s originally 
present at $M_s$ have received a heavy mass and have disappeared from
the low energy spectrum. Thus the SM gauge group is present at low energy.
Also, we have shown that all additional particles 
created from the non-zero intersections of the extra branes
with the colour a-brane and the U(1) leptonic  d, e, f, branes
receive a mass of the order of the string scale and thus are not present 
to low energies. The SM at low energy is really attainable only if we 
show that the fermions  beyond those incorporated in  $F_L$, 
${\bar F}_R$ of table receive a non-zero mass. 
We will now show that this is the case, emphasizing that only the   
weak fermion doublets $\chi_L^1$, $\chi_L^2$, $\chi_L^3$ receive a light mass 
between $M_Z$ and the order of the electroweak symmetry breaking 
$\upsilon$.

Hence the left handed fermions $\chi_L^1$ receive a mass 
of order $\upsilon^2/M_s$ 
from the coupling  
\beq
(1, 2, 1)(1, 2, 1) e^{-A}
 \frac{\langle {\bar h}_3 \rangle \langle {\bar h}_3 \rangle
\langle {F}_R^H  \rangle \langle H_2 \rangle }{M_s^3}
\stackrel{A \rightarrow 0}{\sim}
\frac{\upsilon^2}{M_s} \ (1, 2, 1)(1, 2, 1)
\label{ka1sa1}
\eeq
that is in representation form
\beqa
(1, 2, 1)_{(0, 1, 0, -1, 0, 0)} \ (1, 2, 1)_{(0, 1, 0, -1, 0 0)} 
\langle (1, {\bar 2}, {2})_{(0, -1, 1, 0, 0, 0)} \rangle \
\langle(1, {\bar 2}, {2})_{(0, -1, 1, 0, 0, 0)} \rangle &\nonumber\\
\times \ \langle(4, 1, {\bar 2})_{(1, 0, -1, 0, 0, 0)}\rangle \
\langle({\bar 4}, 1, {\bar 2})_{(-1, 0, -1, 0, 0, 0)}\rangle  
\label{ka1sa111}
\eeqa
In (\ref{ka1sa1}) we have incorporated the leading contribution of the 
worksheet area connecting the
six vertices. 
In the following this contribution will be set for simplicity to
one ($A \rightarrow 0$).

The left handed fermions $\chi_L^2$
receive a mass of order $\upsilon^2/M_s$ 
from the coupling 
\beq
(1, 2, 1)(1, 2, 1) 
 \frac{\langle {\bar h}_3 \rangle \langle {\bar h}_2 \rangle
\langle {F}_R^H  \rangle \langle H_2 \rangle
\langle  {\bar s}_B^2 \rangle}{M_s^4}
\stackrel{A \rightarrow 0}{\sim}
\frac{\upsilon^2}{M_s} \ (1, 2, 1)(1, 2, 1)
\label{ka1sa2}
\eeq
that is in representation form :
\beqa
(1, 2, 1)_{(0, 1, 0, 0, -1, 0)} \ (1, 2, 1)_{(0, 1, 0, 0, -1, 0)} 
\langle (1, {\bar 2}, {2})_{(0, -1, 1, 0, 0, 0)} \rangle \
\langle(1, {\bar 2}, {2})_{(0, -1, 1, 0, 0, 0)}
\rangle &\nonumber\\
\times \ \langle(4, 1, {\bar 2})_{(1, 0, -1, 0, 0, 0)}
\rangle \
\langle({\bar 4}, 1, {\bar 2})_{(-1, 0, -1, 0, 0, 0)}\rangle \ 
\langle (1, 1, 1)_{(0, 0, 0, 0, 2, 0)}\rangle \
\label{ka1sa112}
\eeqa

The $\chi_L^3$ doublet fermions receive heavy masses
of order $\upsilon^2/M_s$ from the realization of the coupling :
\beq
(1, 1, 2)(1, 1, 2)\frac{\langle h_3 \rangle \langle h_3 \rangle
\langle F_R^H \rangle  \langle  H_2 \rangle
\langle {\bar s}_B^3 \rangle}{M_s^3}
\label{real21}
\eeq
In explicit representation form
\beqa
(1, 1, 2)_{(0, 1, 0, 0, 0, -1)}
 \ (1, 1, 2)_{(0, 1, 1, 0, 0, -1)} \
\langle (4, 1, {\bar 2})_{(1, 0, -1, 0, 0, 0)} \rangle \
\langle ({\bar 4}, 1, {\bar 2})_{(-1, 0, -1, 0, 0, 0)} \rangle
\nonumber\\
\times \langle (1, 1, 1)_{(0, 0, 0, 0, 0, 2}\rangle 
\label{real200}
 \eeqa
Thus the left handed fermion weak doublets $\chi_L^1$, $\chi_L^2$, 
$\chi_L^3$    
 receive a low mass of order $\upsilon^2/M_s$. 
 This is a general
 prediction of all classes of GUT models based on non-supersymmetric 
toroidally intersecting D6-branes \cite{tessera2}.
In (\ref{ka1sa1}), (\ref{ka1sa2}), (\ref{real21}) we have 
assumed \footnote{Also assume in the following discussion.} vev's  
$<H_2> \sim <F_R^H>  \sim M_s$. 
For a general string model the issue of determining the size of the
vev's and weather these fields really receive a vev may be made precise  
only after the calculation of the effective potential.

The $\chi_R^1$ right handed doublet fermions receive heavy masses
of order $M_s$ in the following way:
\beq
(1, 1, 2)(1, 1, 2)\frac{\langle H_2 \rangle
\langle F_R^H \rangle 
\langle s_B^1 \rangle}{M_s^2}
\label{real211}
\eeq
In explicit representation form
\beqa
 (1, 1, 2)_{(0, 0, 1, 1, 0, 0)}  \ (1, 1, 2)_{(0, 0, 1, 1, 0, 0)} \
\langle ({\bar 4}, 1, {\bar 2})_{(-1, 0, -1, 0, 0, 0)} \rangle \
\langle (4, 1, {\bar 2})_{(1, 0, -1, 0, 0, 0)} \rangle
\nonumber\\ \times
\langle (1, 1, 1)_{(0, 0, 0, -2, 0, 0}\rangle  
\label{real2001a}
 \eeqa
\newline
The $\chi_R^2$ right handed doublet fermions receive heavy masses
of order $M_s$ from the following coupling
\beq
(1, 1, 2)(1, 1, 2) \frac{\langle H_2 \rangle
\langle F_R^H \rangle 
\langle s_B^2 \rangle}{M_s^2}
\label{real211a}
\eeq
In explicit representation form
\beqa
 (1, 1, 2)_{(0, 0, 1, 0, 1, 0)}  \ (1, 1, 2)_{(0, 0, 1, 0, 1, 0)} \
\langle ({\bar 4}, 1, {\bar 2})_{(-1, 0, -1, 0, 0, 0)} \rangle \
\langle (4, 1, {\bar 2})_{(1, 0, -1, 0, 0, 0)} \rangle
\nonumber\\ \times
\langle (1, 1, 1)_{(0, 0, 0, 0, -2, 0}\rangle  
\label{real200b}
 \eeqa
 \newline                 
The $\chi_R^3$ right handed doublet fermions receive heavy masses
of order $M_s$ in the following way:
\beq
(1, 1, 2)(1, 1, 2)\frac{\langle H_2 \rangle
\langle F_R^H \rangle 
\langle s_B^3 \rangle}{M_s^2}
\label{real2112}
\eeq
In explicit representation form
\beqa
 (1, 1, 2)_{(0, 0, 1, 0, 0, 1)}  \ (1, 1, 2)_{(0, 0, 1, 0, 0, 1)} \
\langle ({\bar 4}, 1, {\bar 2})_{(-1, 0, -1, 0, 0, 0)} \rangle \
\langle (4, 1, {\bar 2})_{(1, 0, -1, 0, 0, 0)} \rangle
\nonumber\\ \times
\langle (1, 1, 1)_{(0, 0, 0, 0, 0, -2}\rangle  
\label{real2001}
 \eeqa

The 6-plet fermions, $\omega_L$, receive a mass term of
order $M_s$  from the
coupling, 
\beq
({\bar 6}, 1, 1)({\bar 6}, 1, 1)
\frac{\langle H_1  \rangle \langle {F}_R^H
\rangle \langle H_1 \rangle \langle {F}_R^H
\rangle}{M_s^3}
\label{6plet}
\eeq
where we have made use of the $SU(4)$ tensor products
$6 \otimes 6 = 1 + 15 + 20$, $ 4 \otimes 4 = 6 + 10$
and have defined
\beqa
({\bar 6}, 1, 1)_{(-2, 0, 0, 0, 0, 0)}
\ ({\bar 6}, 1, 1)_{(-2, 0, 0, 0, 0, 0)}
\langle(4, 1, 2)_{(1, 0, 1, 0, 0, 0)} \rangle \
\langle(4, 1, 2)_{(1, 0, 1, 0, 0, 0)} \rangle &\nonumber\\
\times \ \langle (4, 1, {\bar 2})_{(1, 0, -1, 0, 0, 0)})
\rangle \ \langle(4, 1, {\bar 2})_{(1, 0, -1, 0, 0, 0) })
\rangle
\label{6plet1}
\eeqa

The 10-plet fermions $z_R$ receive
 a heavy mass of order $M_s$ from the coupling
\beq
(10, 1, 1)(10, 1, 1)\frac{\langle {\bar F}_R^H
\rangle \langle {\bar F}_R^H \rangle \langle H_2
\rangle \langle H_2 \rangle}{M_s^3},
\label{10plet}
\eeq
and we have used the
tensor product representations of $SU(4)$,
$10 \otimes 10 = 20 + 35 + 45$,
$20 \otimes {\bar 4} = {\bar 15 } + {\bar 20}$, 
${\bar 20} \otimes {\bar 4} = {\bar 6 } + 10$,
$10 \otimes {\bar 4} =  4  + 36$, $4 \otimes {\bar 4} = 1 + 15$.
Explicitly, in representation form, 
\beqa
(10, 1, 1)_{(2, 0, 0, 0, 0, 0)} (10, 1, 1)_{(2, 0, 0, 0, 0, 0)}
\langle({\bar 4}, 1, 2)_{(-1, 0, 1, 0, 0, 0)}\rangle \
\langle({\bar 4}, 1, {2})_{(-1, 0, 1, 0, 0, 0)}\rangle
&\nonumber\\
\times  \
\langle({\bar 4}, 1, {\bar 2})_{(-1, 0, -1, 0, 0, 0)}\rangle \
\langle({\bar 4}, 1, {\bar 2})_{(-1, 0, -1, 0, 0, 0)}\rangle
\label{10pletagain}
\eeqa
Thus at low energies of order $M_Z$ only the SM remains.

%%%%%%%%%%%%%%%%%%%%%%%%%%%%%%%%%%%%%%%%%%%%
%%%%%%%%%%%%%%%%%%%%%%%%%%%%%%%%%%%%%%%%%%%%%%%%%%%%%%%
%%%%%%%%%%%%%%%%%%%%%%%%%%%%%%%%%%%%%%%%%%%%%%%%%%%%%%%%%%%%%%

\section{Conclusions}

Based on the intersecting D6-brane constructions
of \cite{tessera2},
in \cite{kokos1} we constructed the first examples of string
GUT models
which break at exactly the SM at low energy without any
additional
group factors and/or exotic massless matter.
The classes of GUT models we have been considering
recently \cite{kokos1, kokos2, kokos3} and at the present work, have as
 their low energy theory in
energies of order $M_z$ the Standard model.
Their common characteristic is that they represent
{\em deformations} around the basic intersection
structure of the Quark and Lepton structure,
\beq
I_{ab} = 3, \  I_{ac^{\star}} = -3 \ .
\label{deforme}
\eeq
Thus they all share the same intersection numbers along the
`baryonic' $a$ and the left and right `weak'
$b$ and $c$,  D6 branes.

Interestingly the GUT
constructions have a number of features
independently of the
number of
D6-stacks that they are defined originally. These general
characteristics
include:

\begin{itemize}

\item The prediction of low
mass ($\sim \frac{\upsilon}{M_s}$)
   weak left handed doublets ($\chi_L^i$, $i=1,2,3$)
with mass between $M_Z$ and $246$ GeV. This is a
universal feature and appears at GUT constructions with various
number of
stacks \cite{kokos1} \cite{kokos2} \cite{kokos3}.

This result makes automatic the existence of a low scale 
in the models, e.g. below 650 GeV which makes intersecting 
D6-brane GUTS
directly testable at present or near feature accelerators.

\item The conditions for some of the U(1)'s to survive massless the 
Green-Schwarz mechanism are equivalent to the conditions
originating from the existence
of N=1 supersymmetry in some sectors.
 The preservation of N=1 SUSY is necessary in some
open string sectors in order to allow
 the generation of gauge singlets making massive
 the unwanted
 exotic fermions, e.g.
$\chi_R^1$, $\chi_R^2$, $\chi_R^3$, $s_R^1$, $s_R^2$, $s_R^3$ of 
table (1), and to generate the Majorana mass term.  

\item The presence of extra branes needed to cancel the RR tadpoles, may be 
arranged so that it has 
non-zero intersection numbers with the colour
brane and the rest, but b, c,
of the branes. This feature is to be contrasted with models,
with only the SM at low energy build from a Standard like
configuration at the
string scale $M_s$ contained between four- \cite{louis2}
five- \cite{kokos}
and six- \cite{kokoss} stacks of D6-branes (there is no
configuration of D6-branes able to give only the SM at low
energy beyond 6-stacks),
where the extra branes have no intersection
number with the rest
of the branes. 
The additional fermions created are made always massive by
allowing
the presence of N=1 SUSY in sectors in the
form $jj^{\star}$ and also in sectors having
extra and U(1) leptonic branes.

\item Even though
the models are overall non-supersymmetric they contain
N=1 SUSY preserving sectors. These sectors are necessary
to be implemented into the theory as without them
it will not be possible to generate gauge
singlets breaking the surviving massless the Green-Schwarz
mechanism U(1)'s. Equally important, in their absence
we could not been able to generate a Majorana mass
term for right handed neutrinos.

\end{itemize}

The present non-supersymmetric constructions, if the angle 
stabilization conditions of appendix A hold are free of tachyons, 
however NSNS tadpoles remain, thus leaving the full question of
stability
in these models an open question. We should
nevertheless
remember that even supersymmetric constructions
are free of NSNS tadpoles,
supersymmetry breaking may create an unexpected cosmological constant.  
These tadpoles could be removed in 
principle by background redefinition in
terms of wrapped metrics \cite{du} or by orbifolding as was suggested 
in \cite{34or} and freezing the moduli to discrete values.

However, it is intriguing that we have found intersecting brane 
constructions, in the absence of a dynamical mechanism 
which can select a particular string vacuum, that
offer the possibility to obtain vacua with just the observable 
Standard Model spectrum and gauge interactions at low 
energy.

\begin{center}
{\bf Acknowledgments}
\end{center}
I am grateful
to Luis Ib\'a\~nez
and Angel Uranga
for useful discussions. In addition, I would like to thank the 
Quantum Field Theory group of the Humboldt University of Berlin for its warm 
hospitality, where part of this work was carried out.

\newpage

\section{Appendix I}

In this appendix we list the conditions
  under which the PS-III model classes of
  intersecting D6-branes discussed in this work
  are tachyon free.
Note that 
the conditions are expressed in terms of the angles
defined in (\ref{angPSII}).
We have included the contributions
from the sectors $ab^{\star}$, $ac$, $bd$,
$cd^{\star}$, $be$, $ce^{\star}$, $bf$, $cf{\star}$.
We have not included the
tachyon free conditions from the sectors $dd^{\star}$,
$ee^{\star}$, $ff^{\star}$, as the latter
conditions will
be shown to be
equivalent to the
presence of N=1 supersymmetry in these sectors.

\beqa
\begin{array}{ccccccc}
-(\frac{\pi}{2} + \theta_1) &+& \theta_2 &+& 2 {\theta}_3 &\geq& 0\\
-(\frac{\pi}{2}-{\tilde \theta}_1) &+& \theta_2 &+& 
2{\vartheta}_3 &\geq &0 \\
-(\theta_1 - \frac{\pi}{2}) &+& {\tilde \theta}_2
&+& 2 {\theta}_3 &\geq& 0\\
-(\frac{\pi}{2} + {\tilde \theta}_1) &
+&{\tilde \theta}_2 &+&
2{\theta}_3 &\geq& 0 \\
-( { \theta}_1 -\frac{\pi}{2}  ) &+&{\bar
\theta}_2 &+&
2 {\theta}_3 &\geq& 0 \\
-(\frac{\pi}{2} + {\tilde \theta}_1) &+&{\bar
\theta}_2 &+&
2{\theta}_3 &\geq& 0 \\
-(  {\theta}_1 -\frac{\pi}{2} ) &+&
{\theta}_2^{\prime} &+&
2{\theta}_3 &\geq& 0 \\
-(\frac{\pi}{2} + {\tilde \theta}_1) &+&
{\theta}_2^{\prime} &+&
2{\theta}_3 &\geq& 0 
\\\\
(\frac{\pi}{2} + \theta_1) &-& \theta_2 &+& 2 {\theta}_3 &\geq& 0\\
(\frac{\pi}{2}-{\tilde \theta}_1) &-& \theta_2 &+& 
2{\vartheta}_3 &\geq &0 \\
(\theta_1 - \frac{\pi}{2}) &-& {\tilde \theta}_2
&+& 2 {\theta}_3 &\geq& 0\\
(\frac{\pi}{2} + {\tilde \theta}_1) &-&
{\tilde \theta}_2 &+&
2{\theta}_3 &\geq& 0 \\
( { \theta}_1 -\frac{\pi}{2}  ) &-&{\bar
\theta}_2 &+&
2 {\theta}_3 &\geq& 0 \\
(\frac{\pi}{2} + {\tilde \theta}_1) &-&{\bar
\theta}_2 &+&
2{\theta}_3 &\geq& 0 \\
 ( {\theta}_1 -\frac{\pi}{2} ) &-&
{\theta}_2^{\prime} &+&
2{\theta}_3 &\geq& 0 \\
(\frac{\pi}{2} + {\tilde \theta}_1) &-&
{\theta}_2^{\prime} &+&
2{\theta}_3 &\geq& 0 
\end{array}
\eeqa

\beqa
\begin{array}{ccccccc}
(\frac{\pi}{2} + \theta_1) &+& \theta_2 &-& 2 {\theta}_3 &\geq& 0\\
(\frac{\pi}{2}-{\tilde \theta}_1) &+& \theta_2 &-& 
2{\vartheta}_3 &\geq &0 \\
(\theta_1 - \frac{\pi}{2}) &+& {\tilde \theta}_2
&-& 2 {\theta}_3 &\geq& 0\\
(\frac{\pi}{2} + {\tilde \theta}_1) &+&
{\tilde \theta}_2 &-&
2{\theta}_3 &\geq& 0 \\
( { \theta}_1 -\frac{\pi}{2}  ) &+&{\bar
\theta}_2 &-&
2 {\theta}_3 &\geq& 0 \\
(\frac{\pi}{2} + {\tilde \theta}_1) &+&{\bar
\theta}_2 &-&
2{\theta}_3 &\geq& 0 \\
 ( {\theta}_1 -\frac{\pi}{2} ) &+&
{\theta}_2^{\prime} &-&
2{\theta}_3 &\geq& 0 \\
(\frac{\pi}{2} + {\tilde \theta}_1) &+&
{\theta}_2^{\prime} &-&
2{\theta}_3 &\geq& 0 
\label{free}
\end{array}
\eeqa


\newpage

\begin{thebibliography}{199}


\bibitem{tessera}R.~Blumenhagen, L.~G\"orlich, B.~K\"ors and D.~L\"ust,
``Noncommutative compactifications of type I strings on tori with magnetic
background flux'',
JHEP {\bf 0010} (2000) 006, {\tt hep-th/0007024};
``Magnetic Flux in Toroidal Type I Compactification'',
Fortsch. Phys. 49
(2001) 591, hep-th/0010198 

\bibitem{tessera2}R. Blumenhagen, B. K\"ors and D. L\"ust,
``Type I Strings
with F and B-flux'',
JHEP 0102 (2001) 030, hep-th/0012156.

\bibitem{louis2}
L.~E.~Ib\'a\~nez, F.~Marchesano and R.~Rabad\'an,
``Getting just the standard model at intersecting branes''
JHEP, 0111 (2001) 002, {\tt hep-th/0105155};
L.~E.~Ib\'a\~nez,
``Standard Model Engineering with Intersecting Branes'',
hep-ph/0109082

\bibitem{kokos}C. Kokorelis, ``New Standard Model Vacua
from Intersecting Branes'',
JHEP 09 (2002) 029, 
hep-th/0205147

\bibitem{kokoss}C. Kokorelis,``Exact Standard Model
Compactifications from Intersecting Branes'',
JHEP 08 (2002) 036, hep-th/0206108

\bibitem{kokos1}C. Kokorelis, ''GUT model Hierachies
from
Intersecting Branes'', JHEP 08 (2002) 018,
hep-th/0203187;\\
C. Kokorelis,``Exact Standard Model Structures from
Intersecting
Branes", hep-th/0210004;
C. Kokorelis, Standard Model compactifications from
Intersecting branes, hep-th/0211091


\bibitem{kokos2}C. ~Kokorelis, ``Deformed Intersecting D6-Branes I",
JHEP 2011 (2002) 027, hep-th/0209202

\bibitem{kokos3}C. ~Kokorelis, ``Deformed Intersecting D6-Branes II",
hep-th/0210200;


\bibitem{D5}D. Cremades, L.~E.~Ib\'a\~nez and
 F.~Marchesano,  `Standard model at intersecting
 D5-Branes: lowering the string scale', Nucl.Phys. B643 (2002) 93, 
hep-th/0205074 

\bibitem{D51}C. ~Kokorelis, "Exact Standard model Structures
from Intersecting D5-branes", hep-th/0207234


\bibitem{dikomou}
L. Dixon, V. Kaplunovsky, and J. Louis,
%``Moduli Dependence of String Loop Corrections to Gauge Coupling Constants'',
Nucl. Phys. B355 (1991) 649;\\
P. Mayr and S. Stieberger, `` Threshold Corrections to Gauge Couplings 
in Orbifold Compactifications   '', Nucl. Phys. B407 (1993) 725-748       \\
 C.~Kokorelis, ``String Loop Threshold Corrections for $N=1$
 Generalized Coxeter Orbifolds'',
 Nucl. Phys. B579 (2000) 267, hep-th/0001217;\\
D. Bailin, A. Love, W. Sabra, S. Thomas,  ``String Loop Threshold Corrections
for $Z_N$ Coxeter Orbifolds'', Mod. Phys. Let. A9 (1994) 67, hep-th/9310008;\\
 G.L.Cardoso, D.L\"ust, T. Mohaupt, ``Threshold Corrections and Symmetry Enhancement in String Compactifications'', Nucl.Phys. B450 (1995) 115, 
hep-th/9412209;\\
 C.~Kokorelis, ``Gauge and
 Gravitational Couplings from Modular Orbits in
 Orbifold Compatifications'', Phys. Lett. B477 (2000) 313 ,
 hep-th/0001062;








\bibitem{antoba}
I. Antoniadis, N.Arkadi-Hamed, S. Dimopoulos and G. Dvali,
\PLB436 (1998) 257; I. Antoniadis and C. Bachas, \PLB450 (1999) 83



\bibitem{bele}M. Berkooz, M. R. Douglas, R.G. Leigh, 
``Branes Intersecting at Angles'',
\NPB480 (1996) 265, 
hep-th/9606139

\bibitem{eksi1}
M.~Bianchi, G.~Pradisi and A.~Sagnotti,
``Toroidal compactification and symme
try breaking in open string theories,''
Nucl.\ Phys.\ B376, 365 (1992)



\bibitem{eksi2}
Z.~Kakushadze, G.~Shiu and S.-H.~H.~Tye,
``Type IIB orientifolds with NS-NS antisymmetric tensor backgrounds,''
Phys.\ Rev.\ D58, 086001 (1998).
hep-th/9803141

\bibitem{eksi3}C. Angelantonj,
``Comments on Open-String Orbifolds with a Non-Vanishing 
$B_{ab}$'',   \NPB 566 (2000) 126,   hep-th/9908064

\bibitem{carlo}
R. Blumenhagen, L. G\"orlish, and B. K\"ors, ``Asymmetric Orbifolds, 
non-
commutative geometry and type I string vacua'', 
Nucl. Phys. B582 (2000) 44, hep-th/0003024; 
C. Angelantonj and A. Sagnotti, ``Type I vacua and brane transmutation'', 
hep-th/00010279;
C. Angelantonj, I. Antoniadis, E. Dudas and A. Sagnotti,
``Type I strings on magnetized orbifolds and brane transmutation'',
Phys. Lett. B489 (2000) 223, hep-th/0007090


\bibitem{bachas}C. Bachas, ``A Way to break supersymmetry'',
{\tt hep-th/9503030}.


\bibitem{luis1}
G.~Aldazabal, S.~Franco, L.~E.~Ib\'a\~nez, R.~Rabad\'an and
A.~M.~Uranga,
``D=4 chiral string compactifications from intersecting branes'',
J. Math. Phys. 42 (2001) 3103-3126,
{\tt hep-th/0011073};
G.~Aldazabal, S.~Franco, L.~E.~Ib\'a\~nez, R.~Rabad\'an and
A.~M.~Uranga,
``Intersecting brane worlds'',
JHEP {\bf 0102} (2001) 047, {\tt hep-ph/0011132}.



\bibitem{uran}
R. Blumenhagen, L. G\"orlich, B. K\"ors,
``Supersymmetric Orientifolds in 6D with D-Branes at Angles'',
Nucl.Phys. B569 (2000) 209-228,
hep-th/9908130; 
R. Blumenhagen, L. G\"orlish, B. K\"ors,
``Supersymmetric 4D Orientifolds of Type IIA with D6-branes at
Angles'',
JHEP 0001 (2000) 040, hep-th/9912204;  

\bibitem{fors}S. F\"orste, G. Honecker, R. Schreyer,
``Supersymmetric $Z_N \times Z_M$ Orientifolds in 4D with
D-Branes at Angles'',
Nucl.Phys. B593 (2001) 127, hep-th/0008250;
S. F\"orste, G. Honecker, R. Schreyer, ``Orientifolds with branes at angles'',
JHEP 0106 (2001) 004, hep-th/0105208
G. Honecker, ``Intersecting brane world models from D8-branes on
$(T^2 \times T^4/Z_3)/\Omega {\cal R}_1$ type IIA orientifolds'',
JHEP 0201 (2002) 025, hep-th/0201037; 





\bibitem{cala}R. Blumenhagen, V. Braun, B. K\"ors and D. L\"ust,
``Orientifolds of K3 and Calabi-Yau manifolds with
Intersecting D-Branes'', JHEP 0207 (2002) 026,
hep-th/0206038

\bibitem{uran1}A. M Uranga, ``Local models for intersecting
brane worlds'', hep-th/0208014


%\bibitem{alda} L. F. Alday and G. Aldazabal, ``In quest of 
%"just" the Standard Model on D-branes at a singularity'', 
%hep-th/0203129

\bibitem{cim}D. Cremades,  L.~E.~Ib\'a\~nez, F. Marchesano,
`SUSY Quivers, Intersecting Branes and the Modest Hierarchy Problem',
HEP 0207 (2002) 022, hep-th/0201205; D. Cremades,  L.~E.~Ib\'a\~nez,
F. Marchesano, ''Intersecting Brane Models of
Particle Physics and the Higgs Mechanism'', JHEP 0207 (2002) 009, 
hep-th/0203160; D. Cremades,  L.~E.~Ib\'a\~nez,
F. Marchesano, ``Towards a theory of quark masses, mixings and 
CP-violation'', hep-ph/0212064

\bibitem{cim1}G. Aldazabal, L.E. Ibanez, A. M. Uranga, 
``Gauging Away the Strong CP Problem'', hep-ph/0205250 ;
A. M. Uranga, ``D-brane, fluxes and chirality'', JHEP 0204 (2002) 016,
hep-th/0201221

%\bibitem{nano} J. Ellis, P. Kanti and D. V Nanopoulos
%``Intersecting branes flipped SU(5)'', hep-th/0206087

\bibitem{maria}M. Gomez-Reino and I. Zavala,
``Recombination of Intersecting D-branes and Cosmological
Inflation'', hep-th/0207278

\bibitem{dimi}D. Ghilencea, ``U(1) masses in intersecting 
D-brane SM-like models'', hep-ph/0208205

\bibitem{bkl}D. Bailin, G. V. Kraniotis and A. Love,
`` Intersecting D5-brane models with massive vector-like leptons'',
hep-th/0212112

\bibitem{cve}M. Cvetic, G. Shiu, A. M. Uranga,
``Chiral four dimensional N=1 supersymmetric type IIA orientifolds from
intersecting D6 branes'', 
Nucl. Phys. B615 (2001) 3, hep-th/0107166;
%``Three family
%supersymmetric standard models from intersecting brane worlds'',
%Phys. Rev. Lett. 87 (2001) 201801, hep-th/0107143;
M. Cvetic, P. Langacker, G. Shiu, 
``Phenomenology of A Three-Family Standard-like String Model'',
hep-ph/0205252;
M. Cvetic, P. Langacker, G. Shiu,  
``A Three-Family Standard-like Orientifold model: Yukawa couplings and 
hierarchy'', Nucl. Phys. B 642 (2002) 139, hep-th/0206115;
M. Cvetic, I. Papadimitriou, G.Shiu,  
``Supersymmetric Three Family SU(5) Grand Unified Models 
from Type IIA Orientifolds with Intersecting D6-Branes'', hep-th/0212177 


\bibitem{antokt1}I. Antoniadis, E. Kiritsis 
and T. Tomaras, 
``D-brane Standard Model'' \PLB486 (2000) 186,
hep-th/0111269; I. Antoniadis, E. Kiritsis,
`G. Rizos, T. Tomaras, ``D-branes and the Standard Model'', hep-th/0210263


\bibitem{pati}J. Pati and A. Salam, ``Lepton number as a 
fourth colour'',
Phys. Rev. D10 (1974) 275



\bibitem{antoI}I. Antoniadis, G. K. Leontaris and J. Rizos, ``
A three generation $SU(4) \times O(4)$ string model''
\PLB245 (1990)161;

%\bibitem{antoII}G. K. Leontaris and J. Rizos,
%``A Pati-Salam model from branes'',
% \PLB510 (2001) 295, hep-ph/012255

\bibitem{giapo}A. Murayama and A. Toon,
``An $SU(4) \otimes SU(2)_L \otimes SU(2)_R$ string model
with direct unification at the string scale'',
\PLB318 (1993)298,




%\bibitem{bere} D. Berenstein, V. Jejjala and R.G. Leigh,
% ``The Standard Model on a D-brane'',
%hep-ph/0105042


\bibitem{sa}A. Sagnotti,
``A Note on the Green - Schwarz Mechanism in Open - String Theories'',
 Phys. Lett. B294 (1992) 196, hep-th/9210127



%\bibitem{iru}L.~E.~Ib\'a\~nez, R.~Rabad\'an and A. M. Uranga,
%``Anomalous U(1)'s in Type I and Type IIB D=4, N=1 string vacua'',
%Nucl.Phys. B542 (1999) 112-138;\\
% C. Srucca and M. Serone, ``Gauge and Gravitational anomalies in
%D=4 N=1 orientifolds'', JHEP 9912 (1999) 024


\bibitem{senn}A. Sen, JHEP 9808 (1998) 012, 
``Tachyon Condensation on the Brane Antibrane System''
hep-th/9805170; 


\bibitem{ellis}M. S. Chanowitz, J. Ellis and M. K. Gailard,
``The price of natural flavour conservation in neutral weak interactions'',
\NPB 128 (1977) 506


\bibitem{fi}W. Fischer and L. Susskind, Phys. Lett. B171
(1986) 383; Phys. Lett. B173 (1986) 262



\bibitem{du}
E.~Dudas and J.~Mourad,
``Brane solutions in strings with broken supersymmetry and dilaton
 tadpoles'',
Phys. Lett. B486 (2000) 172, hep-th/0004165;
R.~Blumenhagen, A.~Font,
``Dilaton tadpoles, warped geometries and large extra dimensions for
nonsupersymmetric strings'',
Nucl. Phys. B 599 (2001) 241, {\tt hep-th/0011269}.

\bibitem{34or} R.~Blumenhagen, B.~K\"ors, D.~L\"ust and T. Ott,
``The Standard Model from Stable Intersecting Brane World Orbifolds'',
\NPB616 (2001) 3, hep-th/0107138


 
\end{thebibliography}
\end{document}